\documentclass[11pt]{article}
\pdfoutput=1
\usepackage{jheppub}

\usepackage[latin1]{inputenc}
\usepackage{amsmath}
\usepackage{amsfonts}
\usepackage{amssymb}
\usepackage{graphicx}
\usepackage{xcolor}
\usepackage{mathtools}
\usepackage{simplewick}
\usepackage{hyperref}
\usepackage{multirow}
\usepackage{enumitem}

\newcommand{\Tr}{\text{Tr}}

\begin{document}

\title{
Black holes from large N singlet models
}
\author{Irene Amado$^1$, Bo Sundborg$^1$, Larus Thorlacius$^{1,2}$ and Nico Wintergerst$^{1,3}$}
\affiliation{$^1$The Oskar Klein Centre for Cosmoparticle Physics, Department of Physics, Stockholm University, AlbaNova, 106 91 Stockholm, Sweden}
\affiliation{$^2$University of Iceland, Science Institute, Dunhaga 3, 107 Reykjavik, Iceland}
\affiliation{$^3$The Niels Bohr Institute, University of Copenhagen,
	Blegdamsvej 17, 2100 Copenhagen \O, Denmark}
\emailAdd{irene.amado@fysik.su.se}
\emailAdd{bo@fysik.su.se}
\emailAdd{larus.thorlacius@fysik.su.se}
\emailAdd{nico.wintergerst@nbi.ku.dk}
\abstract{
The emergent nature of spacetime geometry and black holes can be 
directly probed in simple holographic duals of higher 
spin gravity and tensionless string theory. To this end, we study time 
dependent thermal correlation functions of gauge invariant observables 
in suitably chosen free large $N$ gauge theories. At low temperature and 
on short time scales the correlation functions encode propagation through 
an approximate AdS spacetime while interesting departures emerge at 
high temperature and on longer time scales. This includes the existence 
of evanescent modes and the exponential decay of time dependent 
boundary correlations, both of which are well known indicators of bulk 
black holes in AdS/CFT. In addition, a new time scale emerges after which the correlation functions return to a bulk thermal 
AdS form up to an overall temperature dependent normalization. 
A corresponding length scale was seen in equal time correlation 
functions in the same models in our earlier work.
}

\maketitle   
\section{Introduction}
The singlet sector of large $N$ gauge theories coupled to vector or adjoint 
representation scalar matter provides a tractable framework for studying 
gauge/gravity duality in an interesting limit where the 't Hooft coupling 
vanishes. In this limit various explicit computations can be carried out on
the gauge theory side and it is an interesting question to what extent the 
results may be interpreted in terms of an emergent bulk spacetime
on the gravity side. 
We consider two classes of singlet models in various dimensions, 
with matter fields in the vector or adjoint representation of
the gauge group. A representative of each class can be obtained 
as the free limit of an interacting gauge theory arising in well-studied 
examples of gauge/gravity duality. 
Three dimensional Chern-Simons large~$N$ gauge theory with 
matter fields in the vector representation of the gauge group is 
conjectured to be dual \cite{KlebanovPolyakov2002,GiombiMinwallaPrakashTrivediWadiaYin2012,AharonyGur-AriYacoby2012}  to Vasiliev higher spin theory \cite{FradkinVasiliev1987b,FradkinVasiliev1987a} 
while four dimensional Yang-Mills theory with adjoint matter fields is dual to strings 
in an AdS background. The latter is of course the standard 
AdS/CFT duality but in the limit of zero 't Hooft coupling the strings on the
gravity side are tensionless and the dynamics includes an infinite tower of 
higher spin states in addition to the usual massless sector involving the 
graviton and dilaton~\cite{Sundborg2001}. 

A key aspect of gauge/gravity duality is the connection between thermal 
states in the field theory and black holes in the bulk \cite{Witten1998a}. Due to the 
singlet constraint, the free gauge theories considered here exhibit non-trivial 
thermal behavior, including large~$N$ phase transitions at finite volume,
even in the limit of vanishing coupling \cite{Sundborg2000}. In recent work \cite{AmadoSundborgThorlaciusWintergerst2017}, we have
extended the thermodynamic analysis of these models by considering equal 
time correlation functions at thermal equilibrium in the boundary theory and 
analyzing their spatial structure as a function of temperature. 
At low temperature our results are consistent with propagation through 
an AdS bulk geometry with non-trivial additional structure appearing above 
the phase transition. This holds for both adjoint and vector matter, 
but the geometric features that arise at high temperature do so on very different 
length scales in the two types of models. 
For adjoint matter the phase transition is
at the AdS scale and a good case can be made for the black hole interpretation 
by turning on interactions and going to strong coupling where the large~$N$ phase 
transition is transmuted into the Hawking-Page transition and the high temperature 
phase is dominated by bulk black holes \cite{Witten1998a}. 
For vector matter, on the other hand, the phase transition temperature in 
the free theory grows with a power of $N$ compared to the AdS scale in the 
large~$N$ limit \cite{ShenkerYin2011} and any dual bulk objects would appear 
at a characteristic length scale that is much larger than the AdS length \cite{AmadoSundborgThorlaciusWintergerst2017}.

In classical gravity a black hole can be detected by its effect on the 
time evolution of matter in its environment, for instance in-falling particles or waves. 
Boundary correlation functions encode such information in their time dependence.
In the present paper we therefore extend our previous work to the temporal sector of the
singlet models and gather more direct evidence for black hole-like behavior in 
the emergent dual bulk spacetime.  
We map out characteristic features of temporal correlations at different temperatures
and compare them to known black hole signatures in strongly coupled field theories. 
The overall structure is similar in theories with vector and adjoint matter but there
are important differences and we point these out as we go along.    
We note that a closely related problem has been studied earlier in planar AdS/CFT with 
the standard $\mathcal{N}=4$ SYM theory on the boundary \cite{HartnollKumar2005}. 
Having a spherical boundary is essential for our work as it permits us to resolve 
thermal phase transitions and study their influence on correlation functions. 
We can compare our results to the planar case by taking the limit of ultra-high 
temperatures and considering short distances and times relative to the size of 
the spatial sphere. 

We find it useful to consider free singlet theories in general dimensions as toy models
but it is only for the canonical values of three and four dimensions, for theories with 
vector and adjoint matter respectively, that one expects to find interacting conformal 
theories when the gauge coupling is turned on. By keeping the dimension general,
we observe qualitatively different time dependence of the boundary correlation functions 
between even and odd boundary dimensions. This is related to well-known differences 
in wave propagation in even and odd dimensions but seems to have received limited 
attention in the AdS/CFT literature where the main focus has been on spatial correlations 
and on even boundary dimensions. 

Our main results may be listed as follows: 
\begin{enumerate}
\item \label{it:evanescence} At high temperature, the imaginary part of the Fourier space 
retarded propagator receives exponentially suppressed, but non-vanishing, contributions 
at $k \gg T \gg \omega$. This shows the presence of so called evanescent modes which 
are characteristic of an event horizon in a bulk black hole spacetime.  
\item \label{it:depart} 
Boundary correlation functions depart significantly from thermal AdS boundary-to-boundary 
Green's functions at high temperature, as should be the case in the presence of a localized 
central object in the bulk.  
	\begin{enumerate}
	\item \label{it:exponential} In even boundary dimensions, they undergo a period of 
	rapid exponential decay, as is known to occur in an AdS black hole background in 
	Einstein gravity.
	\item \label{it:ripple} In odd boundary dimensions, the retarded Green's function develops 
	support inside the light cone. 
	\end{enumerate}
\item \label{it:int_scale} For low boundary dimensions, high-temperature correlators 
exhibit a characteristic time scale beyond which their form returns to that of thermal 
AdS correlators with a temperature dependent normalization factor. This feature, 
suggesting further localized structure, is also seen in higher boundary dimensions, 
but then only over a finite range of temperatures above the critical temperature.
\end{enumerate}

The paper is planned along the following lines.
Section~\ref{sec:phase_transition} contains some background on the models
under study and their large $N$ phase transitions. This is well known material 
that we include to establish notation and set the stage for what follows.  
In Section~\ref{sec:time_dependence} we express the time dependent correlation 
function of a pair of singlet operators at finite 
temperature in terms of the distribution of eigenvalue of the
$U(N)$ gauge holonomy around the thermal $S^1$. 
The resulting formulas are analyzed in the remainder of the paper.  
We first focus on properties of the retarded propagator in Section~\ref{sec:evanescence}
and find that evanescent modes (item \ref{it:evanescence} above) are present in the 
high-temperature phase with adjoint matter, while for vector matter they are already 
discernible below the phase transition at temperatures 
that are parametrically comparable to the critical temperature. 
The existence of such modes has been argued previously from a collective field theory 
approach to thermo-field dynamics of the boundary theory in a high-temperature limit, 
well above the phase transition \cite{JevickiSuzuki2016,JevickiYoon2016}. Our results 
provide a direct confirmation of evanescent modes in singlet models at any temperature
that is sufficiently high for the eigenvalue distribution to deviate from a flat distribution.
In Section~\ref{sec:general_behavior} we turn to a comprehensive analysis of the 
time-ordered correlator, mainly using analytic methods but also presenting results 
obtained by numerical means. We establish a period of exponential decay in time of 
correlations in even boundary dimensions and show that the retarded Green's function 
is non-zero in the interior of the light cone in odd dimensions (as per item \ref{it:depart} 
above). For intermediate temperatures, a detailed numerical study is presented in 
Section~\ref{sec:intermediate}. In particular, we point out features that a naive 
high-temperature expansion fails to capture. Those features are then the focus of 
Section~\ref{sec:non-classical}, where we catch a glimpse of intriguing finite volume
behavior in the high temperature phase (item \ref{it:int_scale}). Beyond a certain 
timescale, the boundary correlation functions return to the form of thermal 
boundary-to-boundary propagators in an AdS bulk geometry, dressed by a 
temperature dependent normalization factor. 
We wrap up with a discussion in Section~\ref{sec:discussion} and this is followed 
by several appendices with additional technical details and background material.

\section{Eigenvalue distributions and large-$N$ phase transitions}
\label{sec:phase_transition}

One of our main goals is to calculate time dependent correlation functions 
in $U(N)$ gauge theories in the limit of vanishing gauge coupling
with a view towards an interpretation in terms of emergent 
spacetime geometry in a dual gravitational theory. We consider 
matter in the form of a free scalar field in the adjoint representation 
of the gauge group or a finite number $N_f$ of scalars in the
fundamental representation. We work at finite volume and finite 
temperature with fields defined on a spatial unit $d{-}1$--sphere 
and a thermal circle $t\cong t+i\beta$.

In free gauge theories, extracting nontrivial thermodynamic features is a 
matter of counting singlet operators, for example by Polya's method, as done 
in the original work \cite{Sundborg2000}, where the projection to singlets was also introduced
by hand as a matrix model for a $U(N)$ valued Lagrange 
multiplier. Alternatively, one can derive the matrix integral from the 
path integral expression for the partition 
function \cite{AharonyMarsanoMinwallaPapadodimasVan-Raamsdonk2004} for 
interacting gauge theory. By viewing the singlet 
constraint as the residual Gauss law of the gauge theory at zero coupling, 
the Lagrange multiplier has a straightforward interpretation as the $U(N)$ 
gauge holonomy around the thermal circle.
 
At large~$N$, the matrix model can be solved in a saddle point approximation. 
This is achieved by approximating the integral over the eigenvalues of the 
matrix valued Lagrange multiplier as a path integral over an eigenvalue 
distribution $\rho(\lambda)$. Extremizing the corresponding action leads to a 
saddle point expression for $\rho(\lambda)$, which in turn governs the physical
behavior of correlation functions. 
Derivations of $\rho(\lambda)$ for matter in the adjoint 
and fundamental representations can be found in \cite{GrossWitten1980,Sundborg2000,AharonyMarsanoMinwallaPapadodimasVan-Raamsdonk2004,ShenkerYin2011} and we summarize the main
steps involved in Appendix~\ref{sec:ev_dists}. 

\subsection{Vector model}
The saddle point value of the eigenvalue density in a theory with $N_f$ 
scalar fields in the fundamental representation of $U(N)$ is \cite{GrossWitten1980}
\begin{equation}
	\label{eq:dist_sum}
	\rho(\lambda) = \frac{1}{2\pi} 
	+ \frac{N_f}{N}\sum_{k=1}^{\infty} z_S^d(x^k) \frac{1}{\pi}\cos(k\lambda)\,,
\end{equation}
where $-\pi\leq\lambda\leq\pi$ and $z_S^d(x)$ is the one-particle partition sum 
on a $d-1$-dimensional sphere,
\begin{equation}
	z_S^d(x)=x^{\frac{d}{2} - 1}\frac{1+x}{(1-x)^{d-1}}\,,
\end{equation}
with $x = e^{-\beta}$. Here, and in the following, we have chosen units in which the 
sphere has unit radius and $\beta = 1/T$ is dimensionless.

The eigenvalue density is constrained to be positive and satisfies the normalization 
condition $\int d\lambda\,\rho(\lambda) = 1$. Low temperature
corresponds to $x\ll 1$ and it immediately follows from \eqref{eq:dist_sum} that  
the eigenvalue distribution is flat,
\begin{equation}
\rho(\lambda)\approx \frac{1}{2\pi},
\end{equation}
in the limit of low temperature and large~$N$. 
At high temperature, on the other hand,
we have $x\rightarrow 1$. In this case the one-particle partition sum simplifies
and the eigenvalue density reduces to
\begin{equation}
	\label{eq:dist_highT}
	\rho(\lambda) = \frac{1}{2\pi} 
	+ \frac{2N_f }{\pi N}\,T^{d-1}\sum_{k=1}^{\infty} \frac{\cos(k\lambda)}{k^{d-1}}\,.
\end{equation}
This expression is valid as long as $\rho(\lambda)$ remains everywhere positive 
but this fails at a critical temperature,
\begin{equation}
\label{Tcritical}
T_c^{d-1} = \frac{1}{4\left(1-2^{2-d}\right)\zeta(d-1)}\, \frac{N}{N_f}\,,
\end{equation}
at which $\rho(\pm\pi)\rightarrow 0$. The critical temperature $T_c$, which is parametrically 
high in this model, signals a third order phase transition~\cite{GrossWitten1980}.

\begin{figure}[t]
	\centering{
		\includegraphics[width=.49\textwidth]{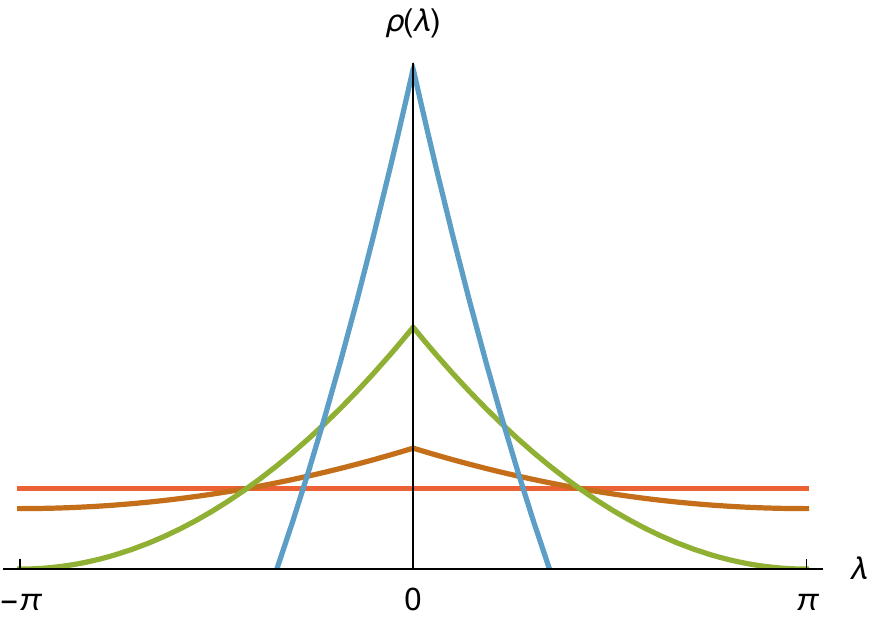}\hfill		
		\includegraphics[width=.49\textwidth]{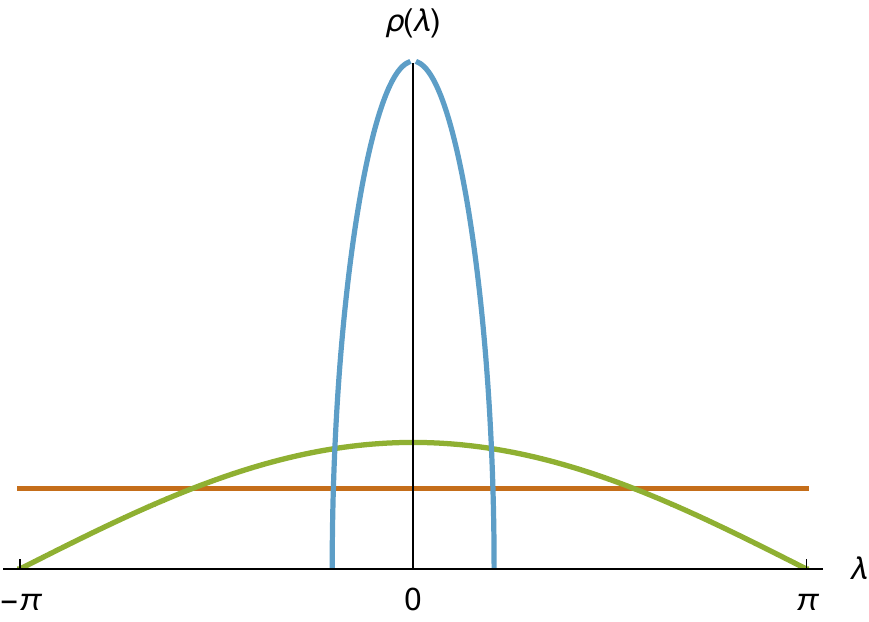}
		\caption{\label{fig:evdists} \emph{Left:} Eigenvalue distribution in 
		vector model at $d=3$ for $T/T_c \ll 1$ (red), $T/T_c = 1/2$ (orange), 
		$T/T_c = 1$ (green) and $T/T_c \approx 2$ (blue). 
		\emph{Right:} Eigenvalue distribution in adjoint model at $d=4$ for 
		$T/T_c \ll 1$ (red), $T/T_c = 1/2$ (orange), $T/T_c = 1$ (green) 
		and $T/T_c \approx 2$ (blue). In this case, curve remains the 
		same for all $T\leq T_c$ and the phase transition is first order. 
		For $T/T_c \to \infty$, the distribution approaches a $\delta$-function in both models.
		}
	}
\end{figure}

The shape of the eigenvalue distribution at different temperatures is shown in 
Fig.~\ref{fig:evdists}. Above the phase transition the eigenvalue 
density vanishes beyond a certain maximal value $\lambda_m$ determined
by the normalization condition $\int_{-\lambda_m}^{\lambda_m}\rho(\lambda)d\lambda=1$.
Explicit formulas are given in Appendix~\ref{sec:ev_dists} for the 
eigenvalue density at high temperature as well 
as for the maximal eigenvalue $\lambda_m$ above the phase transition. 
At very high temperatures the eigenvalue density becomes narrowly 
peaked around the origin and approaches a Dirac delta function in
the $T\rightarrow\infty$ limit. For later reference, we here quote the scaling formulas for the width of the distribution at high 
temperature, $T\gg T_c$, in different spatial dimensions, as derived in Appendix~\ref{sec:ev_dists},
\begin{equation}\label{eq:lambdam}
\lambda_m \to \frac{T_c}{T} \times
\begin{cases}
\sqrt{\frac{2}{\pi A_d}} &\text{for }d=3\,,\\
\left(\frac{3}{2A_d}\right)^{\frac{1}{3}}\log^{-1/3}\frac{T}{T_c}&\text{for }d=4\,,\\
\left(\frac{3}{2\zeta(d-3)A_d}\right)^{\frac{1}{3}}
\left(\frac{T_c}{T}\right)^{\frac{d-4}{3}} &\text{for }d\geq 5\,.
\end{cases}
\end{equation}
In particular, $\lambda_m\sim T_{c}/T$ in $d=3$, the canonical dimension for the vector model. 

\subsection{Adjoint model}
The thermodynamic behavior is similar when the matter field is in the adjoint 
representation of the gauge group but there are important differences. In particular, the
critical temperature remains finite in the large~$N$ limit and the phase transition is first order. 
The eigenvalue distribution is shown in the right hand panel in Figure~\ref{fig:evdists}. 
It remains flat, $\rho(\lambda) = \frac{1}{2\pi}$, at all temperatures below the phase transition
but above the transition the eigenvalue density develops a non-trivial profile that 
can be approximated by the following 
expression~\cite{AharonyMarsanoMinwallaPapadodimasVan-Raamsdonk2004},
\begin{equation}
\label{approxrho}
\rho(\lambda) = 
\begin{cases}
\frac{1}{\pi \sin^2\left(\frac{\lambda_m}{2}\right)}\sqrt{\sin^2\left(\frac{\lambda_m}{2}\right) 
- \sin^2\left(\frac{\lambda}{2}\right)}\cos\left(\frac{\lambda}{2}\right)&\text{if} \quad
0\leq|\lambda| \leq \lambda_m\,, \\
0&\text{if} \quad \lambda_m<|\lambda| \leq \pi\,.
\end{cases}
\end{equation}
The maximal eigenvalue $\lambda_m$ is determined by
\begin{equation}
\sin^2\left(\frac{\lambda_m}{2}\right) = 1 - \sqrt{1-\frac{1}{z_S^d(x)}}\,.
\end{equation}
At high temperature, $T\gg T_c$, the eigenvalue distribution of the adjoint model 
becomes narrowly peaked,
\begin{equation}
\rho(\lambda) \approx \frac{2}{\pi \lambda_m}\sqrt{1 - \frac{\lambda^2}{\lambda_m^2}}\,,
\end{equation}
with the maximal eigenvalue going to zero as an inverse power of temperature,
$\lambda_m \rightarrow T^\frac{1-d}{2}$.

\section{Time dependent correlators}\label{sec:time_dependence}

In this section, we present the main equations of our paper. We begin with the 
general expression for the time ordered thermal correlator, from which all subsequent 
results can be derived. We then specialize to the retarded Green's function.
The central quantity of interest is the time-ordered correlator of two gauge-singlet operators.
In the vector model it reads
\begin{align}
G(t,{\bf y}) &\equiv \frac{2^{d+2}\pi^{d}}{\Gamma^2(\frac{d-2}{2})} 
\frac{1}{N}\left\langle \Tr |\Phi^2(t,{\bf y})| \Tr |\Phi^2(0)| \right\rangle \nonumber \\&= 
\sum_{j=1}^N \left(\sum_{m=-\infty}^{\infty}  \frac{e^{i m \lambda_j}}{(\cos(t+i\beta m) 
-  \cos\theta)^\frac{d-2}{2}}\right)^2\nonumber\\
&=\int_{-\pi}^{\pi} d\lambda\,\rho(\lambda)\left(\sum_{m=-\infty}^{\infty}  
\frac{e^{i m \lambda}}{(\cos(t+i\beta m) -  \cos\theta)^\frac{d-2}{2}}\right)^2
\,,
\label{eq:singsingreallamvec}
\end{align}
where we have introduced a multiplicative constant on the first line that cancels the 
constant numerical factor that arises from the Wick contractions of the scalars. This 
has no effect on the physics but streamlines all subsequent equations. The explicit factor
of $1/N$ guarantees that the two-point correlation function does not scale with $N$.

The corresponding expression in the adjoint model is
\begin{align}
G(t,{\bf y}) &\equiv 
\frac{2^{d+2}\pi^{d}}{\Gamma^2(\frac{d-2}{2})} 
\frac{1}{N^2}\left\langle \Tr |\Phi^2(t,{\bf y})| \Tr |\Phi^2(0)| \right\rangle \nonumber \\
&=\int_{-\pi}^{\pi} d\lambda\,d\lambda'\,\rho(\lambda)\rho(\lambda')
\left(\sum_{m=-\infty}^{\infty}  \frac{e^{i m (\lambda-\lambda')}}{(\cos(t+i\beta m) 
-  \cos\theta)^\frac{d-2}{2}}\right)^2
\,.
\label{eq:singsingreallamadj}
\end{align}
The final equality in either equation holds in the continuum limit $N \to \infty$ and the 
dependence of Eq.~\eqref{eq:singsingreallamadj} on the difference of eigenvalues is a 
manifestation of the unbroken center symmetry of the underlying gauge 
theory \cite{AharonyMarsanoMinwallaPapadodimasVan-Raamsdonk2004}.
We observe that the correlation function in the adjoint theory can be obtained from 
the one in the vector theory by simply replacing $\lambda$ by $\lambda - \lambda'$ 
in \eqref{eq:singsingreallamvec} and including an additional integral\footnote{Note that 
by explicitly limiting the support of the eigenvalue distribution to $(-\pi ,\pi]$ and extending 
the domain of integration to the whole real line, we can alternatively express 
Eq.~\eqref{eq:singsingreallamadj} as 
	\begin{equation}
	\int_{-\infty}^{\infty} d\lambda\,(\rho\star\rho)(\lambda)
	\left(\sum_{m=-\infty}^{\infty}  \frac{e^{i m \lambda}}{(\cos(t+i\beta m) 
	-  \cos\theta)^\frac{d-2}{2}}\right)^2\,.\nonumber
	\end{equation}
Adjoint model results are then obtained by replacing the eigenvalue distribution 
$\rho(\lambda)$ by the convolution $\rho\star\rho$.} over $\lambda'$. 
As a consequence, we will mainly show derivations for the vector model in the following, 
and only display final results for the adjoint theory.

The above expressions for time dependent correlators are most readily obtained through analytic 
continuation of the equal time correlators of \cite{AmadoSundborgThorlaciusWintergerst2017},
\begin{equation}
G({\bf y}) =
\sum_j\,
\left(\sum_{m} \frac{e^{i m \lambda_j}}{\left(\cosh m\beta 
	- \cos\theta\right)^\sigma}\right)^2
\end{equation}
by sending $\beta m \to \beta m - i t$. The denominator follows from the corresponding
expression for the vacuum two point correlator of the same singlet operators.
The nontrivial new ingredient here compared to the vacuum result is the dependence 
on the eigenvalues of the $U(N)$ matrix that implements the singlet constraint. 
For completeness, we include a 
derivation from first principles in Appendix \ref{app:greenfcn}.\footnote{Following \cite{AmadoSundborgThorlaciusWintergerst2017}, one can perform the integral over 
$\lambda$ to obtain an expression in terms of Fourier cosine coefficients of the 
eigenvalue distribution. Since we will not make much use of that representation in the
present work, we refrain from showing it here.}
Note that we have effectively implemented a thermal normal ordering procedure that 
subtracts the square of the one-point function from the correlation functions. In a different 
language, we have redefined the field operators such that their one-point function vanishes.

In what follows, we will also be using the retarded correlator of singlet operators, defined as
\begin{equation}
G_R(x,t) \equiv \frac{2^{d+2}\pi^{d}}{\Gamma^2(\frac{d-2}{2})} \frac{1}{N} 
\Theta(t)\, \text{Im}\left\langle \Tr |\Phi^2(x,t)| \Tr |\Phi^2(0)| \right\rangle \,.
\end{equation}
In order to extract it from Eq.~\eqref{eq:singsingreallamvec}, it is useful to rewrite the 
sum over $n$ inside the integral over eigenvalues,
\begin{equation}
G(t,{\bf x})= \int_{-\pi}^{\pi} d\lambda\,\rho(\lambda)
\left(\frac{1}{(\cos t -  \cos\theta)^\frac{d-2}{2}}
+2\sum_{m=1}^{\infty} \text{Re} \left[\frac{e^{i m \lambda}}{(\cos(t+i\beta m) 
-  \cos\theta)^\frac{d-2}{2}}\right]\right)^2\,.
\end{equation}
Since the thermal sums are manifestly real and non-singular, any imaginary contribution 
must necessarily involve the vacuum piece. Consequently, we have
\begin{multline}
G_R(x,t) = \Theta(t)\Bigg\{\text{Im}\left[\frac{1}{(\cos t -  \cos\theta)^{d-2}}\right]\\
+4\,\text{Im}\left[\frac{1}{(\cos t - \cos\theta)^{\frac{d-2}{2}}}\right]
\int_{-\pi}^{\pi} d\lambda\,\rho(\lambda)\sum_{m=1}^{\infty} 
\text{Re}\left[ \frac{e^{i m \lambda}}{(\cos(t+i\beta m) -  \cos\theta)^\frac{d-2}{2}}\right]\Bigg\}\,.
\label{eq:gret}
\end{multline}
The first term is the well-known vacuum contribution to the retarded propagator. It is 
insensitive to the singlet constraint and has support only on the boundary light cone. 
The second term is a thermal contribution that depends on the eigenvalue distribution. 
We will discuss it in more detail in section \ref{sec:retarded}. 
In Fourier space, the retarded correlator contains information about the spectrum of 
the theory, which has important implications for the dual bulk physics.
We discuss this next.

\section{Evanescent modes}\label{sec:evanescence}

Wave propagation in an AdS black hole background in standard gravity theory 
includes so called evanescent modes with an exponential fall-off towards the 
boundary whose existence is tied to the event horizon in the bulk spacetime 
geometry \cite{ReyRosenhaus2014}. 
Finding such modes in an exotic theory with massless higher spin fields 
is a strong indication of the presence of black hole-like structures in that theory. 
In this section we demonstrate that the retarded propagator at high temperatures in 
our free singlet models indeed has precisely the right form to support evanescent 
modes in a dual bulk interpretation.

In thermal field theory, the spectral density is defined as the imaginary part of 
the retarded propagator in frequency-momentum space. 
In the low energy phase of singlet theories, it is expected to only receive contributions 
from $\omega^2 \geq {\bf k}^2$. This follows from the interpretation of the low temperature 
phase in terms of a thermal gas of higher spin particles. In the bulk, it receives only on-shell 
contributions and since the square of the momentum in the direction normal to the boundary 
is positive definite, one has $\omega^2 = {\bf k}^2+k_z^2 \geq {\bf k}^2$.

A nontrivial geometric structure in the bulk can, however, support  
modes for which $k_z$ is complex. In this case the mode amplitude is exponentially 
suppressed radially towards the boundary or, conversely, exponentially growing in the 
radial direction into the bulk interior. Such a mode can nevertheless remain normalizable 
if the bulk geometry is such that the exponential growth is cut off at some radial position.
This happens for instance in the presence of a horizon in the bulk 
where modes with complex $k_z$ propagate parallel to the horizon but are 
exponentially suppressed towards the boundary and thus effectively confined to 
the near horizon region. In the corresponding 
boundary theory the signature of a bulk evanescent mode is an exponentially suppressed 
but non-vanishing spectral density in the far off-shell regime $\omega \ll T \ll |{\bf k}|$. 
Such contributions were discovered previously in a planar limit of a free gauge singlet 
model \cite{JevickiYoon2016}. The planar limit corresponds to the infinite temperature 
limit in our conventions. Using our approach, it is straightforward to look for evanescent 
mode signatures at finite temperature and we indeed find that they persist at high
temperatures of both the adjoint and vector models. More specifically, we see that
the presence of evanescent modes is directly linked to having a nontrivial eigenvalue 
distribution in the dual gauge theory. 

As always in the free boundary theories that we are considering, the bulk interpretation
of the results involves a degree of extrapolation. The above picture of evanescent modes 
propagating in the near horizon region of a black hole relies on having a classical bulk 
geometry and thus an effective field theory on the gravitational side of the duality. On the 
other hand, a spectral density of the above form is a robust signal that one can look for in 
any field theory. Accordingly, we can take evanescent behaviour in the boundary spectral 
function as one of the defining features of a bulk black hole away from the limit where we 
have an effective field theory description on the bulk side of the duality. 

Note that since we are ultimately interested in large momenta $k \equiv |{\bf k}| \gg 1$, 
we can use a flat Fourier space representation of the correlation function. This simplifies 
the calculation considerably but it should be kept in mind that the description breaks 
down for small momenta. 

In principle, the spectral density in the evanescent regime could be obtained from a direct 
Fourier transformation of Eq.~\eqref{eq:gret} in the flat limit. We find it convenient, however,
to take an extra step and first obtain the retarded correlator in mixed $k$-$t$-space. 
To this end we use the relation $G_R(k,t) = \Theta(t)\,\text{Im}\,G(k,t)$,
which remains valid in momentum space as the transformation from position space 
does not introduce a complex phase.
The time ordered correlator in $k$-$t$ space can be written as
\begin{equation}
G(k,t) = \int d\lambda \,\rho(\lambda) \int d^{d-1}p_1 \,d^{d-1}p_2 \, 
\delta^{(d-1)}({\bf k} - {\bf p_1} -{\bf p_2})\, G_T(p_1,t,\lambda)G_T(p_2,t,\lambda)
\end{equation}
with 
\begin{equation}
G_{T}(p,t,\lambda) = \frac{1}{2 p} 
\left[e^{-i p t} \left(\Theta(t)+ n_B(p - i\frac{\lambda}{\beta})\right) 
+ e^{i p t} \left(\Theta(-t)+ n_B(p + i\frac{\lambda}{\beta})\right)\right]\,.
\end{equation}
This essentially corresponds to Eq.~\eqref{eq:gf_t_vec_k}, derived in Appendix~\ref{app:greenfcn}. 
In comparison, we have here performed the contraction on the eigenvectors of the gauge 
holonomy, used that $E \approx p$ in the high momentum limit, and replaced the sum over 
eigenvalues by its continuum approximation.
The corresponding retarded correlator can be expressed as
\begin{equation}
\label{eq:gret_mom}
G_R(k,\omega) = \int dt\, d^{d-1}p_1 \,d^{d-1}p_2 \,e^{i \omega t} \delta^{(d-1)}({\bf k} 
- {\bf p_1} -{\bf p_2})\, G_R(p_1,p_2,t)\,,
\end{equation}
with
\begin{align}
G_R(p_1,p_2,t) &= \Theta(t)\,\text{Im}\left[\int d\lambda \,\rho(\lambda)
G_T(p_1,t,\lambda)G_T(p_2,t,\lambda)\right]\nonumber\\&
=\frac{i\Theta(t)}{8 p_1 p_2}\int d\lambda\,\rho(\lambda)
\bigg[\left(1 + n(p_1 + i \frac{\lambda}{\beta})
+n(p_2 + i \frac{\lambda}{\beta})\right)\left(e^{-i t (p_1+p_2)}-e^{i t (p_1+p_2))}\right)\nonumber\\
&\quad+\left(n(p_1 + i \frac{\lambda}{\beta})
-n(p_2 + i \frac{\lambda}{\beta})\right)
\left(e^{it(p_1-p_2)}-e^{-it(p_1-p_2)}\right)
\bigg]\,.
\end{align}
We have used the fact that $\rho(\lambda)$ is even and that there are no purely thermal 
contributions to the retarded correlator in accordance with Eq.~\eqref{eq:gret}. 
Since $G_R(p_1,p_2,t)$ is manifestly real, taking the imaginary part of 
Eq.~\eqref{eq:gret_mom} allows us to drop the $\Theta$-function. 
Integrating over $t$ then yields a set of $\delta$-functions and we immediately 
observe that those that depend on $p_1+p_2$ will not contribute in the evanescent 
regime $\omega \ll k$, due to the $\delta$-function in Eq.~\eqref{eq:gret_mom}. 
Setting them to zero and integrating over $p_2$ in the remaining terms yields 
\begin{multline}
\text{Im}\,G_R(k,\omega) = \int d\lambda\,\rho(\lambda)\int d^{d-1}p\, \frac{1}{8 p |{\bf k}-{\bf p}|}
\left(n(p + i \frac{\lambda}{\beta})-n(|{\bf k}-{\bf p}| + i \frac{\lambda}{\beta})\right)\\
\times
\left(\delta(\omega+p-|{\bf k}-{\bf p}|)-\delta(\omega-p+|{\bf k}-{\bf p}|)\right)
\,.
\end{multline}
By virtue of the $\delta$-functions, we can rewrite this as
\begin{multline}
\text{Im}\,G_R(k,\omega) = \int d\lambda\,\rho(\lambda)\int d^{d-1}p \frac{1}{16p}\\\Bigg[\frac{1}{ p+\omega}
\left(n(p + i \frac{\lambda}{\beta})-n(p+\omega + i \frac{\lambda}{\beta})\right)
\delta(\omega+p-|{\bf k}-{\bf p}|)\\
-\frac{1}{p-\omega}\left(n(p + i \frac{\lambda}{\beta})-n(p-\omega 
+ i \frac{\lambda}{\beta})\right)\delta(\omega-p+|{\bf k}-{\bf p}|)\Bigg]
\,.
\end{multline}
Working to lowest order in $\omega$ in the evanescent regime, this can in turn be approximated by
\begin{equation}
\text{Im}\,G_R(k,\omega) = -\omega\int d\lambda\,\rho(\lambda)\int d^{d-1}p\frac{1}{8 p^2} 
n'(p + i \frac{\lambda}{\beta})\delta(p-|{\bf k}-{\bf p}|)
\,.
\end{equation}
The above expression only has support for $p \gg \beta$. There, the Bose distribution becomes a Boltzmann distribution, allowing us to integrate over $\lambda$ to obtain
\begin{equation}
\text{Im}\,G_R(k,\omega) = \frac{\beta\omega\rho_1}{4 k}\frac{\pi^{\frac{d-2}{2}}}{\Gamma(\frac{d-2}{2})}\int dp\,p^{d-4} d\theta \sin^{d-4}\theta\, e^{-\beta p}\delta(\theta - \arccos\frac{k}{2p})
\,.
\end{equation}
Here, we have assumed a symmetric eigenvalue distribution and introduced its first Fourier cosine moment, $\rho_1 \equiv \int d\lambda\,\rho(\lambda) \cos\lambda$, non-vanishing only for a non-constant eigenvalue distribution.
The limits for the integration over $p$ have to be adjusted such that the $\delta$-function can be satisfied. In the evanescent regime, the limits are $k/2$ and $\infty$.
Integrating over both $\theta$ and $p$ then yields
\begin{align}
\text{Im}\, G_R(k,\omega) &\approx 
\frac{\pi\omega\rho_1}{ 4}\left(\frac{\pi k}{\beta}\right)^{\frac{d-5}{2}}K_{\frac{d-3}{2}}\left(\frac{k\beta}{2}\right)\nonumber\\
&\approx 
\frac{\pi\omega\rho_1}{ 4k}\left(\frac{\pi k}{\beta}\right)^{\frac{d-4}{2}}e^{-\frac{k\beta}{2}}\,.
\label{eq:im_prop_ev_fin}
\end{align}
with the modified Bessel function $K_\alpha(x)$.
Evanescent behavior signaled by the exponential is thus seen above the phase transition, for 
both adjoint and vector models in any number of dimensions $d \geq 3$. In fact, for vector models 
the evanescent modes actually appear already below the phase transition (at temperatures scaling 
as $T_{c}$ with $N$). This is a direct consequence of the phase transition being third order and the Fourier cosine mode $\rho_1$ changing continuously at $T_c$. Similar effects are observed for example in the free energy \cite{ShenkerYin2011}.

We find it remarkable that our simple boundary theories give rise to
black hole-like behavior in the dual higher spin gravity or tensionless string theory. We note that 
the coefficient $\beta/2$ in the exponent that we find in free singlet theories differs from the 
corresponding coefficient obtained in supergravity \cite{SonStarinets2002}. While there is no {\it a priori} reason for the exponents in 
	supergravity and higher spin gravity to agree, it would be 
	interesting to understand the relation between the two.

\section{General behavior of two-point correlation functions}
\label{sec:general_behavior}

In this section we study the time dependence of the correlation function 
Eq.~\eqref{eq:singsingreallamvec} at general temperatures with an eye towards a bulk
interpretation. In particular, we will observe exponentially decaying correlation functions in 
the high temperature phase, a well-known signature of bulk black holes in holographic 
CFTs \cite{Maldacena2003}. Another important result of this section is the emergence of a 
previously unnoticed characteristic time scale at high temperature, beyond which the 
time dependence of the two-point correlation function resembles that of its low 
temperature counterpart. We observed a corresponding length scale in our earlier study 
of equal time correlation functions \cite{AmadoSundborgThorlaciusWintergerst2017}. 

To simplify the analysis, we mainly focus on the so-called autocorrelator, where the 
operators are inserted at coincident spatial points, leaving only a dependence on their 
time difference. The discussion is divided into a few subsections. First, we develop a 
more transparent representation of the correlation function that holds at coincident spatial 
points. We then develop analytic approximations valid at very low and very high 
temperatures respectively. In particular, we contrast ubiquitous power law tails at low 
temperatures with exponential decay in even boundary dimensions in the high 
temperature phase. Following this, we present a numerical study of the correlation
function at intermediate temperatures. Among other things, we uncover a new characteristic 
time scale in the numerical data that appears at high temperature and signals the return to 
a power law decay with time of the autocorrelator. 

\subsection{Coincident points}
\label{sec:coincident}
As before, the derivation we present is for the vector model and only results are quoted 
for adjoint representation scalars. To get started, we use basic trigonometric identities 
(taking the principal value of the square root for odd $d$) to rewrite 
Eq.~\eqref{eq:singsingreallamvec} for $\theta = 0$ as
\begin{equation}
\label{eq:reallamapp}
G(t)  = (-2)^{2-d}
\int d\lambda\,\rho(\lambda)\left(\sum_{m=-\infty}^{\infty} 
\frac{e^{i m \lambda}(\text{sgn}(m))^{d}}{\sin^{d-2}\left(\frac{t+i\beta m}{2}\right)}\right)^2\,,
\end{equation}
where we have assumed $t>0$ for simplicity.
Eq.~\eqref{eq:reallamapp} takes form of an integral over a square of a Fourier sum. 
The discontinuity of the function inside the sum, when viewed as a function of $m$, leads to
complications in odd boundary dimensions that hamper a derivation of a simpler expression 
for arbitrary temperatures. As a result, our results for odd dimensions explicitly will rely on an
approximate evaluation of Eq.~\eqref{eq:reallamapp}. 
In even dimensions, on the other hand, the function is continuous and we can express it 
as a sum over images using the Poisson summation formula,
\begin{equation}
\label{eq:reallampoiss}
G(t)  = (-2)^{2-d}
\int d\lambda\,\rho(\lambda)\left(\sum_{k=-\infty}^{\infty} h_t^{(d)} (\lambda + 2 \pi k) \right)^2  \,,
\end{equation}
where we have defined
\begin{equation}
\label{eq:htdef}
h_t^{(d)}(\lambda) \equiv \int_{-\infty}^{\infty} dm\,e^{i m \lambda} 
\frac{1}{\sin^{d-2}\left(\frac{t+i\beta m}{2}\right)}
\end{equation}
as the Fourier transform of the function inside the summation. The higher dimensional
$h_t^{(d)}(\lambda)$ can be obtained from the $d=4$ expression as follows
\begin{equation}
h_t^{(d)}(\lambda) = \frac{2^{d-4}}{(d-3)!}\left[\prod_{n=1}^{\frac{d-4}{2}}(n^2 
+\partial_t^2)\right]h_t^{(4)}(\lambda)\,,
\end{equation}
so it suffices to calculate the Fourier transform only for $d=4$. 
We obtain
\begin{equation}
\label{eq:ftsum}
h_t^{(4)}(\lambda) = -\frac{4\pi}{\beta} 
\text{csch}\left(\frac{\pi\lambda}{\beta}\right) \partial_t e^{\frac{(\pi - t) \lambda}{\beta }}\,,
\end{equation}
which readily follows from expressing $h_t^{(4)}(\lambda)$ in terms of an incomplete Euler beta 
function, $B_x(a,b) \equiv \int_{0}^{x} s^{a-1}(1-s)^{b-1} ds$ and using the reflection 
identity 
\begin{equation}
	\label{eq:betaref}
	B_z(a,b) = (-z)^{-a}z^a\left(\left(\frac{1}{z}\right)^{a+b}(-z)^{a+b}
	B_{\frac{1}{z}}(1-a-b,b)+B(a,1-a-b)\right)\,.
\end{equation}

The above expression for the autocorrelator is not amenable to analytic arguments 
for general temperatures but it can be evaluated numerically to good precision, as we shall 
see later on in this section. Before presenting our numerical results, however, we want to 
take advantage of certain simplifications that occur at high temperature to obtain 
analytic results that reveal properties of the high-temperature phase of the model.
We have to evaluate the sum
\begin{equation}
\sum_{k=-\infty}^{\infty} e^{-\frac{(\lambda+2\pi k)  t}{\beta }}\,
\left(1+\coth\left(\frac{\pi(\lambda+2\pi k)}{\beta}\right)\right)\,.\label{eq:periodic}
\end{equation}
In the regime $\beta \ll 1$, we can to leading order set the hyperbolic cotangent to 
$\text{sgn}(k)$ unless $k = 0$. This limit holds for $-\pi <\lambda<\pi$, and since 
Eq.~\eqref{eq:periodic} is periodic, the approximation can be periodically extended 
beyond the original region of validity. Contributions for $k < 0$ can then be neglected 
and we obtain for the sum
\begin{equation}
e^{-\frac{\lambda  t}{\beta }}\,\left(1+\coth\left(\frac{\pi\lambda}{\beta}\right)
+2\sum_{k=1}^{\infty} e^{-\frac{2\pi k t}{\beta }}\right)
= e^{-\frac{\lambda  t}{\beta }}\,\left(\coth\left(\frac{\pi t}{\beta}\right)
+\coth\left(\frac{\pi\lambda}{\beta}\right)\right)\,,
\end{equation}
and for the correlator Eq.~\eqref{eq:reallampoiss} in even dimensions for $\beta \ll 1$
\begin{multline}
\label{eq:reallampoisseval}
G(t)  = \frac{2^{d-2}\pi^2}{((d-3)!)^2}\frac{1}{\beta^2} \\
\int d\lambda\,\rho(\lambda)\, \left[\left(\prod_{n=1}^{\frac{d-4}{2}}\left(n^2
+\partial_t^2\right)\right)\partial_t\left( e^{-\frac{\lambda  t}{\beta }}\,
\left(\coth\left(\frac{\pi t}{\beta}\right)+\coth\left(\frac{\pi\lambda}{\beta}\right)\right)\right)\right]^2  \,. 
\end{multline}
For adjoint scalars, Eq.~\eqref{eq:reallampoiss} is replaced by
\begin{equation}
\label{eq:reallampoissadj}
G(t)  = (-2)^{2-d}
\int d\lambda\,d\lambda'\,\rho(\lambda)\rho(\lambda')\left(\sum_{k=-\infty}^{\infty} 
h_t (\lambda-\lambda' + 2 \pi k) \right)^2  \,,
\end{equation}
and the even-dimensional high temperature correlator 
\begin{multline}
\label{eq:reallampoissevaladj}
G(t)  = \frac{2^{d-2}\pi^2}{((d-3)!)^2}\frac{1}{\beta^2} \int d\lambda\,d\lambda'\,
\rho(\lambda)\rho(\lambda')\,\\
\times\left[\left(\prod_{n=1}^{\frac{d-4}{2}}\left(n^2+\partial_t^2\right)\right)\partial_t
\left( e^{-\frac{(\lambda-\lambda')  t}{\beta }}\,\left(\coth\left(\frac{\pi t}{\beta}\right)
+\coth\left(\frac{\pi(\lambda-\lambda')}{\beta}\right)\right)\right)\right]^2 
\end{multline}
takes the place of Eq.~\eqref{eq:reallampoisseval}.

\subsection{Low temperatures: Power laws}
\label{sec:low}

At temperatures below the phase transition in the adjoint case and at $\beta \leq \mathcal{O}(1)$ 
in the vector model, the correlator is obtained from a constant $\rho(\lambda)$. 
The $\lambda$-integration in Eq.~\eqref{eq:reallamapp} then picks out the diagonal 
elements of the square and one obtains
\begin{equation}
G(t)  = 2^{d-2}
\sum_{m=-\infty}^{\infty} \left(\cosh(m\beta)-\cos t\right)^{2-d}\,.
\label{eq:lowt}
\end{equation}
For $t \ll \beta$, the vacuum piece, $m=0$, dominates, implying a power law decay 
for small $t$,
\begin{equation}\label{eq:earlyLowTpower}
G(t) \sim  \frac{1}{t^{2(d-2)}}\,.
\end{equation}

At somewhat higher temperatures, such that there exist times that are much longer 
than the time that corresponds to the thermal wavelength, yet smaller than the AdS 
period, we can consider the leading order contribution in $\beta$. 
In the regime $\beta \ll t \ll 1$ we obtain
\begin{align}
G(t)	&\approx 2^{d-1}\sum_{n=1}^{\infty} \frac{1}{(t^2 + \beta^2 n^2)^{d-2}}\nonumber\\
&\approx
\frac{2^{d-1}}{\Gamma(d-2)}\partial_{t^2}^{d-3}\left[\frac{1}{2t^2}\left(\frac{\pi t}{\beta} 
\coth(\frac{\pi t}{\beta}) - 1\right)\right] \,.
\label{eq:gflow_coinc_flat}
\end{align}
In this regime the autocorrelator approaches another power law tail with an altered exponent
\begin{equation}\label{eq:lateLowTpower}
G(t)	\sim  \frac{1}{\beta t^{2d-5}}\,,
\end{equation}
as could already have been anticipated on dimensional grounds.
The behavior outlined in this subsection persists at all temperatures below the phase
transition, $T < T_c$. We shall demonstrate this towards the end of 
section~\ref{sec:non-classical}, after Eq.~\eqref{eq:ren_power}.

\subsection{Ultra-high temperatures}\label{sec:highest_temperature}
We now consider very high temperatures under the assumption that the eigenvalue 
distribution can be well described by a $\delta$-function.\footnote{As explained in 
Section~\ref{sec:non-classical}, for the vector model this is only valid for $d\geq 4$. However, 
even in the $d=3$ case it proves instructive to compare to the $\delta$-function limit, which is 
why we nevertheless include this case here.}. 
For simplicity, we first focus on insertions at the same point in space, {\it i.e.} set 
$\theta = 0$, and then take $\theta$ to be non-vanishing but small. Moreover, we 
observe rather different behavior of correlation functions in even and odd boundary 
dimensions and thus find it useful to distinguish between the two. 

\subsubsection{Even dimensions: Exponential decay}
\label{sec:exp_decay}

In even boundary dimensions correlation functions exhibit a period of exponential decay in the high temperature phase, 
as expected if black hole-like objects were present in the bulk. We demonstrate this first 
by direct summation of Eq.~\eqref{eq:reallamapp} in the short time limit, where the 
expressions are particularly simple. We then establish the general results that follow from 
Eqs.~\eqref{eq:reallampoisseval} and \eqref{eq:reallampoissevaladj}.

For short times and $\rho(\lambda) = \delta(\lambda)$, we expand the summand in 
Eq.~\eqref{eq:reallamapp} in $t$ and $\beta$, 
\begin{align}
\sum_{m=-\infty}^{\infty} \frac{(\text{sgn}(m))^{d-2}}{\sin^{d-2}\left(\frac{t+i\beta m}{2}\right)}	
&\approx 2^{d-2}\sum_{m=-\infty}^{\infty} \frac{1}{\left(t+i\beta m\right)^{d-2}}\nonumber\\
&\approx
\frac{2^{d-2} \pi}{\beta}\frac{1}{(d-3)!}\frac{\partial^{d-3}}{\partial t^{d-3}}
\coth\left(\frac{\pi t}{\beta}\right)\,.
\label{eq:plan_even}
\end{align}
For the time ordered correlator we then obtain
\begin{equation}
\label{eq:evendplanar}
G(t)	\approx \frac{2^{d-2} \pi^2}{\beta^2}\left[\frac{1}{(d-3)!}
\frac{\partial^{d-3}}{\partial t^{d-3}}\coth\left(\frac{\pi t}{\beta}\right)\right]^2\,.
\end{equation}
For instance, in $d = 4$, we have
\begin{equation}
G(t)	\approx \frac{4\pi^4}{\beta^4}\text{csch}^4(\frac{\pi t}{\beta})\,.
\end{equation}
As opposed to the expression below the transition (see Eqs.~\eqref{eq:earlyLowTpower} 
and \eqref{eq:lateLowTpower} above), no power law tails are present. 
At sufficiently late times, $t \gtrsim \beta$, we thus observe an exponentially 
decaying correlation function in even dimensions.

For arbitrary times, we insert $\rho(\lambda) = \delta(\lambda)$ directly into 
Eqs.~\eqref{eq:reallampoisseval} and \eqref{eq:reallampoissevaladj} for the vector and
adjoint models, respectively. In both cases we obtain
\begin{equation}
\label{eq:reallampoissevalexp}
G(t)  = \frac{2^{d-2}}{((d-3)!)^2}\frac{1}{\beta^2}
 \left[\left(\prod_{n=1}^{\frac{d-4}{2}}\left(n^2+\partial_t^2\right)\right)
 \left( 1+\frac{\pi^2}{\beta}\text{csch}^2\left(\frac{\pi t}{\beta}\right)\right)\right]^2  \,.
\end{equation}
Compared to Eq.~\eqref{eq:evendplanar}, we observe additional contributions that are a 
consequence of considering the theory on a sphere. In particular, we always find a constant 
contribution to the correlators. The exponential decay thus comes to a halt once the 
$t$-dependent part becomes of the same order as this constant piece, at $t \sim \log{T}/T$. 
After that the correlator remains approximately constant for a while. 

We emphasize that this behavior persists for smeared operators. In fact, we can 
obtain all the above results for general $\theta$. Focusing on the simpler example of 
small times, small $\theta$ yields
\begin{align}
G(t,\theta)&\approx 
\frac{2^{d-2}}{\Gamma^2(\frac{d-2}{2})} \left[\partial_{\theta^2}^{\frac{d-4}{2}}
\sum_{n=-\infty}^{\infty} \frac{1}{(\beta n - i t)^2+\theta^2}\right]^2\nonumber\\
&\approx \frac{2^{d-4}\pi^2}{\Gamma^2(\frac{d-2}{2})\beta^2}
\left[\partial_{\theta^2}^{\frac{d-4}{2}}\frac{\coth(\frac{\pi(t+\theta)}{\beta})
-\coth(\frac{\pi(t-\theta)}{\beta})}{\theta}\right]^2 \,.
\label{eq:gfhigh_theta_flat}
\end{align}
Expanding the above for $\theta \ll t$ reveals an exponential decay in time for all angles 
that are well localized inside the light cone. Accordingly, smearing the fields does not 
affect decay properties of the correlation functions in even dimensions. Let us highlight that the exponential decay is a direct consequence of the high temperature phase. We have demonstrated it in the limit of ultrahigh temperatures, and will confirm it at intermediate temperatures $T \gtrsim T_c$ in our numerical analysis in section~\ref{sec:intermediate}, as well as analytically in section~\ref{sec:non-classical}.

\subsubsection{Odd dimensions: Retarded Green's functions supported inside the light cone}
\label{sec:retarded}
As mentioned earlier, away from the light cone, the retarded propagator 
Eq.~\eqref{eq:gret} receives contributions only from the thermal piece, and only in 
odd dimensions. There, it reads
\begin{multline}
G_R(x,t)|_{\cos t \neq \cos\theta} = 4(-1)^{\frac{d-3}{2}}\Theta(t)\Theta(\cos \theta - \cos t)\,
|\cos t - \cos\theta|^{-\frac{d-2}{2}}\\\times\int_{-\pi}^{\pi} d\lambda\,\rho(\lambda)
\sum_{m=1}^{\infty} \text{Re}\left[ \frac{e^{i m \lambda}}{(\cos(t+i\beta m) 
-  \cos\theta)^\frac{d-2}{2}}\right]\,
\end{multline}
for the vector model, while the result in the adjoint model is obtained by inserting the 
corresponding convolution.
As we can see, its nontrivial nature within the light cone is directly linked to a non-constant 
eigenvalue distribution. In Appendix \ref{sec:supportConsistency} we check how absence of 
light cone interior support in even boundary dimensions is consistent with the expected 
support inside the light cone for odd dimensional bulk wave equations.

In the limit of coincident spatial points, we can extract the analytical form of the time-ordered 
correlator from the $\lambda \to 0$ limit of Eq.~\eqref{eq:reallamapp}, by using well known 
properties of the Polygamma functions that appear in an explicit summation. As anticipated, 
we will observe a non-vanishing imaginary part. 
For odd $d$, Eq.~\eqref{eq:reallamapp} becomes 
\begin{align}
G(t) &= -2^{2-d}
\left(\frac{1}{\sin^{d-2}\left(\frac{t}{2}\right)}+2i\sum_{m=1}^{\infty}\text{Im}
\left[ \frac{1}{\sin^{d-2}\left(\frac{t+i\beta m}{2}\right)}\right]\right)^2\nonumber\\
&=
-2^{2-d}\left[\frac{1}{\sin^{d-2}\left(\frac{t}{2}\right)}+\frac{2^{d-2}i}{(d-3)!}
\prod_{n=1}^{\frac{d-3}{2}}\left(\frac{(2n-1)^2}{4}+\partial_t^2\right)
\sum_{m=1}^{\infty}\text{Im}\left[ \frac{1}{\sin\left(\frac{t+i\beta m}{2}\right)}\right]\right]^2
\label{eq:oddht}
\,,
\end{align}
irrespective of the scalar representation.
The last sum essentially corresponds to the definition of the $Q$-Polygamma function 
$\psi_q(z)$ and can be written as 
\begin{equation}
\sum_{m=1}^{\infty}\text{Im}\,\left[ \frac{1}{\sin\left(\frac{t+i\beta m}{2}\right)}\right]
=\frac{2}{\beta}\,\text{Re}\left[\psi_{e^{-\frac{\beta}{2} }}\left(1-i\frac{t}{\beta}\right)
-\psi_{e^{-\frac{\beta}{2} }}\left(1+i\frac{2\pi-t}{\beta}\right)\right]\,.
\end{equation}
For $t>\beta$ the above can be approximated by 
\begin{equation}
\frac{2}{\beta}\,\text{Re}\left[\psi_{e^{-\frac{\beta}{2} }}
\left(1-i\frac{t}{\beta}\right)-\psi_{e^{-\frac{\beta}{2} }}
\left(1+i\frac{2\pi-t}{\beta}\right)\right] \approx -\frac{4}{\beta} \log\cot\frac{t}{4}\,.
\end{equation}
Consequently, one finds for $d = 3$
\begin{equation}
G(t)	\approx -\frac{1}{2}\left[\frac{1}{\sin \frac{t}{2}}-\frac{4i}{\beta} \log\cot\frac{t}{4}\right]^2.
\end{equation}
We note the dimensionally reduced form of the second term, dominant at $t \gg \beta$. 
It is logarithmic at small $t$, just as a massless correlator in $d=2$, i.e.\ in one dimension less. 
This holds in any dimension, since the result in higher dimensions follows by simple 
differentiation. A small $t$ expansion for $d>3$ yields
\begin{equation}
	G(t)	\sim \frac{1}{\beta^2} \frac{1}{t^{2(d-3)}}\,.
\end{equation}
This effective dimensional reduction is a well known feature of thermal quantum field theories. We note in particular the difference in exponent to Eq.\eqref{eq:lateLowTpower}, the late time behavior in the low temperature phase.

The retarded autocorrelator can be read off directly from Eq.~\eqref{eq:oddht}. 
For our example of $d = 3$, one obtains
\begin{equation}
G_R(t)	\approx -\frac{4}{\beta}\frac{\log\cot\frac{t}{4}}{\sin \frac{t}{2}}\,.
\end{equation}
In summary, we observe a nontrivial imaginary part for the retarded correlator inside the 
light cone in odd boundary dimensions in the high temperature phase. Moreover, we have 
given an explicit expression for the time ordered Green's function at spatial coincidence, 
recognizing a dimensionally reduced form for $t \gg \beta$, when thermal effects are dominant.

\subsection{Intermediate temperatures: Numerical results} \label{sec:intermediate}
In an intermediate regime, where the eigenvalue distribution can neither be approximated by 
a constant nor a $\delta$-function, the correlation functions are best evaluated numerically. 
This allows us to single out effects of a nontrivial eigenvalue distribution, 
revealing shortcomings of the $\delta$-function approximation discussed in 
Section~\ref{sec:highest_temperature} in low boundary dimensions. A new time scale $t_{a}$ emerges in our plots.
This is very similar to what we found in equal time correlators in 
\cite{AmadoSundborgThorlaciusWintergerst2017}.  The new scale is explained analytically in Section~\ref{sec:non-classical}.
Since the $\delta$-function essentially cancels the $\lambda$ integration that implements 
the singlet constraint, we refer to the results of the $\delta$-function approximation as the 
\emph{unconstrained theory}. It is included in the plots for comparison.

At sufficiently low temperatures, a direct numerical summation of Eq.~\eqref{eq:reallamapp} 
is feasible. Good precision is obtained for an upper summation limit of the order of $10/\beta$. 
At high temperatures, instead, with $\beta \ll 1$, the expressions to be evaluated are different 
in odd and even boundary dimensions.
In odd dimensions, we interpret Eq.~\eqref{eq:reallamapp} as a Riemann sum, thus effectively 
considering
\begin{equation}
\label{eq:oddlamriem}
G(t)  = (-2)^{2-d}
\int d\lambda\,\rho(\lambda)\left(\frac{1}{\sin^{d-2}\left(\frac{t}{2}\right)}
+2i\,\text{Im}\int_{1}^{\infty} dm\frac{e^{i m \lambda}}{\sin^{d-2}
\left(\frac{t+i\beta m}{2}\right)}\right)^2\,,
\end{equation}
or the adjoint model equivalent thereof.
In even dimensions we instead work directly with 
Eqs.~\eqref{eq:reallampoisseval} and \eqref{eq:reallampoissevaladj}.

Key features of the correlation functions are displayed in the figures.
In discussing their behavior, we distinguish the cases of even and odd dimensions.
All plots were obtained for $N = 10^{5}$ and varying $\beta$. 
In Figs. \ref{fig:fun3dreal} and \ref{fig:fun3dret} we show numerical solutions for both 
the vector and adjoint models for $d=3$. This thus includes the canonical dimension for 
higher spin/CFT duality. We observe the following behavior:

\begin{figure}[t]
	\centering{
		\includegraphics[width=.49\textwidth]{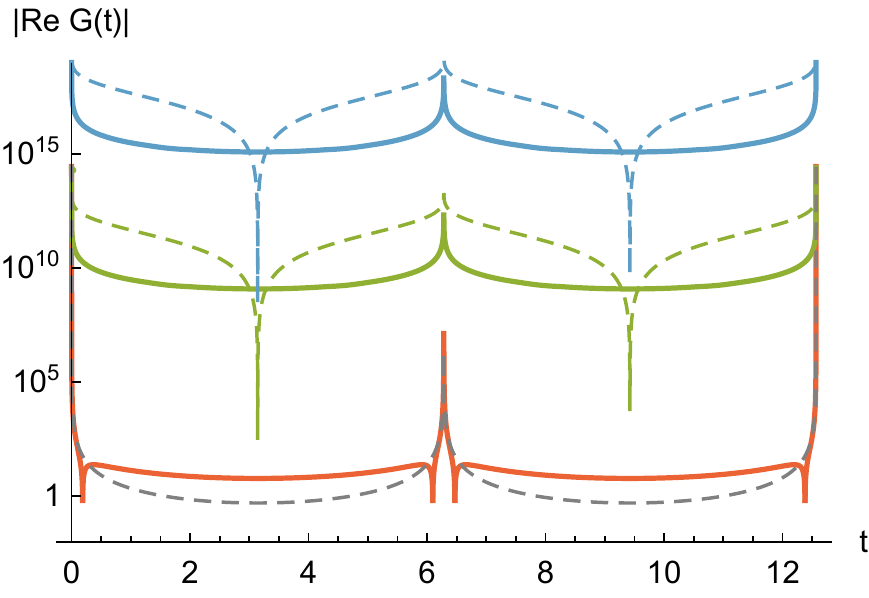}\hfill		\includegraphics[width=.49\textwidth]{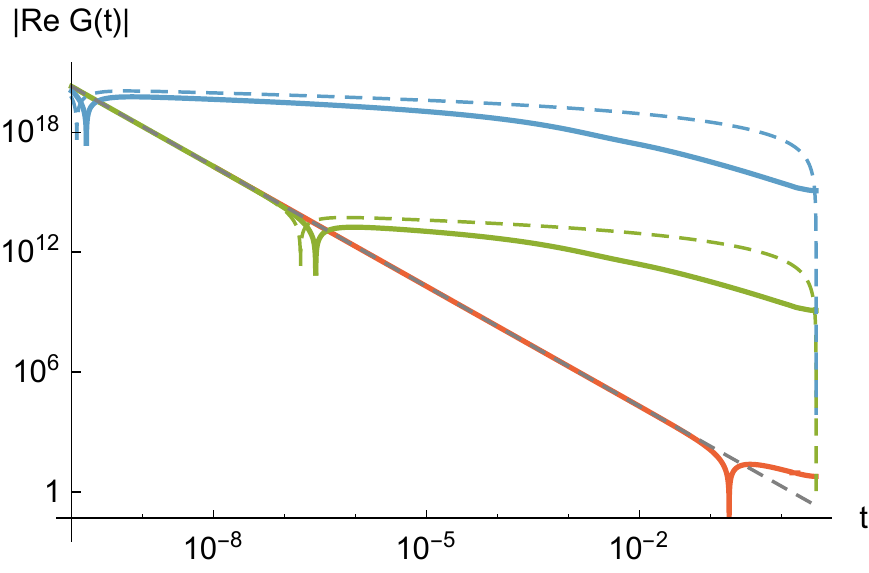}
		\caption{\label{fig:fun3dreal} Vector model, d=3. $|G(t)|$ for $\beta/\beta_c = 50$ (red), $\beta/\beta_c = 10^{-3}$ (green) and $\beta/\beta_c = 10^{-6}$ (blue), with $t_a \sim 3\cdot10^{-3}$ in both high temperature cases. The gray dashed line depicts the vacuum correlator, while the colored dashed lines correspond to an unconstrained theory at the respective temperature, displayed only for $\beta < \beta_c$. The logarithmic plot on the left illustrates the periodicity of the correlators, as well as the renormalized low temperature behavior (shifted red curve) for $t > t_a$. More of the features can be extracted from the log-log plot on the right. 
		}
	}
\end{figure}

\begin{figure}[t]
	\centering{
		\includegraphics[width=.49\textwidth]{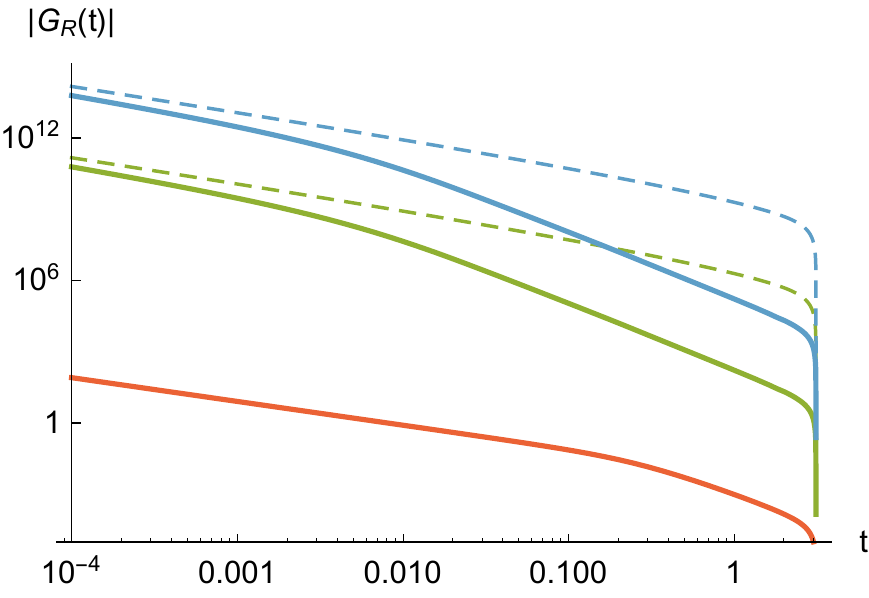}\hfill		\includegraphics[width=.49\textwidth]{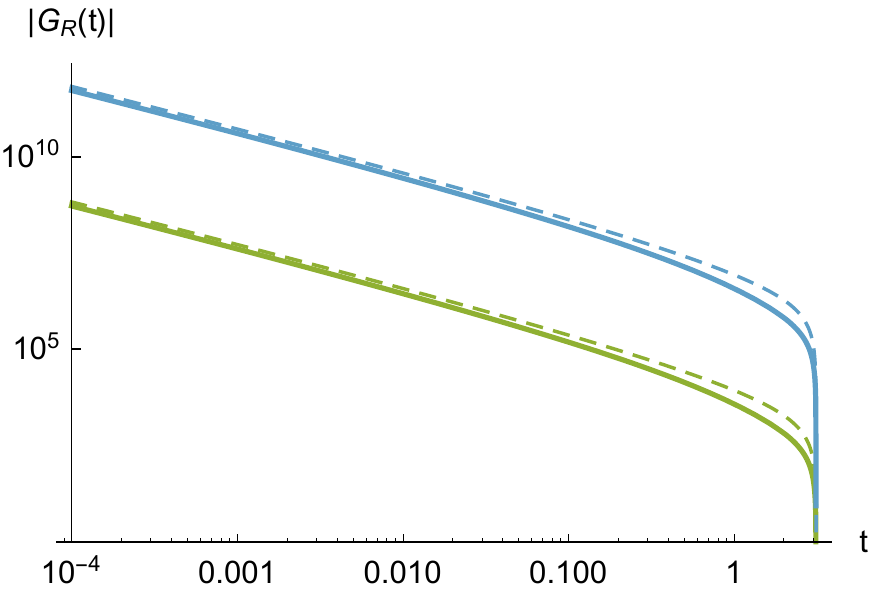}
		\caption{\label{fig:fun3dret} d=3, retarded correlator $G_R(t) = \text{Im }G(t)$. \emph{Left:} Vector model for $\beta/\beta_c = 50$ (red), $\beta/\beta_c = 10^{-3}$ (green) and $\beta/\beta_c = 10^{-6}$ (blue). The dashed lines depict the result of an unconstrained theory at the corresponding temperatures. The crossover times are as in Fig.~\ref{fig:fun3dreal}. \emph{Right:} Adjoint model for same values of $\beta/\beta_c$. The imaginary part vanishes below the transition. The crossover times are at $t_a \sim 1$ for both lines.
		}
	}
\end{figure}

\begin{description}
\item[$t \ll \beta$:] At short times, the real part of the correlator is dominated by the 
	power law decay of the vacuum contribution, included in all plots as a gray dashed line.
\item[$t \sim \beta$:] Characteristic zeros appear. 
	Analogous zeros appear in Green's functions in Einstein gravity black holes in four 
	bulk dimensions, and are associated with ``ringing'' behavior 
	\cite{ChanMann1997,ChanMann1999,CardosoKhanna2015}. In our case, the number 
	of zeros is small and the ringing thus rudimentary. 
\item[$\beta \ll t \ll t_a$:] The decay slows down significantly. For decreasing $\beta/\beta_c$, 
	the correlator approaches that of an unconstrained theory and significantly departs from a 
	thermal AdS boundary-boundary correlation function. We observe an effective dimensional 
	reduction of the correlators, as pointed out towards the end of section \ref{sec:retarded}.
\item[$t \sim t_a$:] At this time, the correlator exhibits a cross-over from the unconstrained 
behavior to a renormalized low temperature behavior, as evident in the left panel of 
Fig.~\ref{fig:fun3dreal}. In $d=3$, this effect prevents the zero-crossing for $t \to \pi$ 
found in the unconstrained theory. In higher dimensions, the crossover scale depends on 
$\beta/\beta_c$ and for sufficiently small $\beta/\beta_c$ increases beyond the radius of 
the sphere \cite{AmadoSundborgThorlaciusWintergerst2017}. Also the retarded propagator 
$G_R(t) = \text{Im }G(t)$, displayed in Fig.~\ref{fig:fun3dret}, is sensitive to this scale. 
Beyond it, it decays significantly faster than its unconstrained counterpart. 
\end{description}

\begin{figure}[t]
	\centering{
		\includegraphics[width=.49\textwidth]{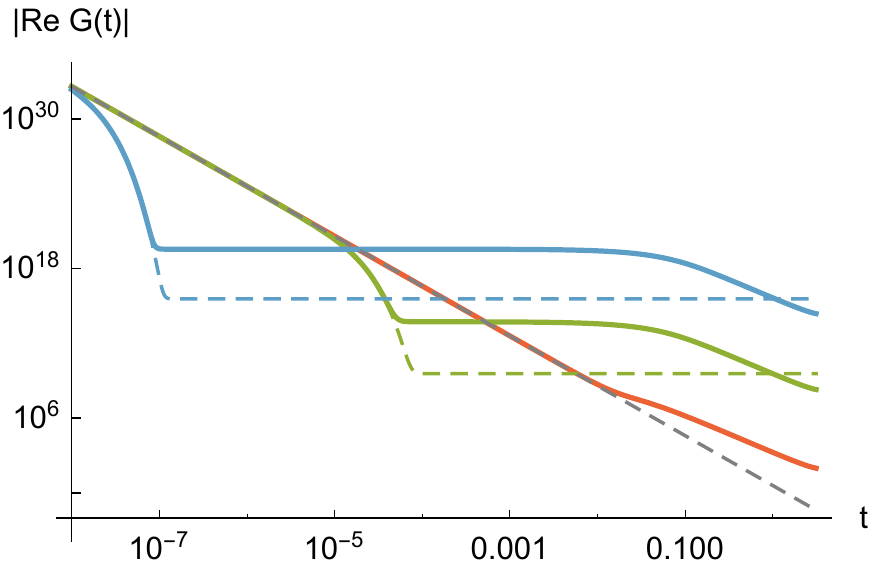}\hfill		\includegraphics[width=.49\textwidth]{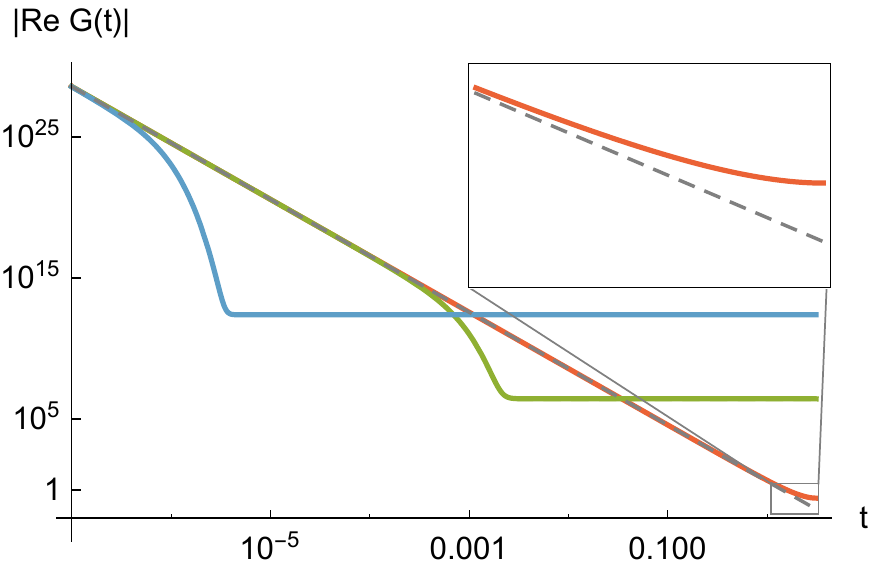}
		\caption{ \label{fig:decay} $d=4$. \emph{Left}: Vector model for $\beta/\beta_c = 10$ (red), $\beta/\beta_c = 10^{-3}$ (green) and $\beta/\beta_c = 10^{-6}$ (blue). The solid lines correspond to the exact behavior, while the dashed lines illustrate the unconstrained theory. The gray dashed line depicts the vacuum correlator. The crossover times in the high temperature phase are in order of increasing temperature $t_a \simeq 0.15$ and $0.18$. \emph{Right}: Adjoint model for the same values of $\beta/\beta_c$. Line styles are as on the left. The crossover times are beyond the light crossing time of the sphere.
		}
	}
\end{figure}

As discussed at length in previous sections, the even dimensional correlators exhibit a 
characteristic exponential decay at times of the order of $\beta$. This can be clearly seen 
in Figure~\ref{fig:decay}, both for fundamental and adjoint matter. In particular, the right 
hand side of the figure corresponds to the zero 't Hooft coupling limit of the well-known 
duality between strings on $AdS_5 \times S^5$ and ${\cal N} = 4$ SYM. We can 
summarize the behavior as follows:
\begin{description}
\item[$t \ll \beta$:] At short times, the correlator is again dominated by the power law 
decay of the vacuum contribution, included in the plots as a gray dashed line.
\item[$t \sim \beta$:] At temperatures above the transition, correlations decay exponentially 
fast. The decay continues until finite volume and finite temperature effects become important.
\item[$\beta \ll t \ll t_a$:] The correlator approaches a constant value set by the temperature 
and by $\lambda_m$. For the vector model, we observe a significant discrepancy between the 
exact behavior and the unconstrained theory, due to the finite width of the eigenvalue 
distribution, manifest in the ratio $\lambda_m/\beta \gtrsim 1$. These subtleties are 
discussed in Section \ref{sec:non-classical}.
\item[$t \sim t_a$:] Around this time the form of the correlations return to that of the 
low-temperature phase provided $\beta/\beta_c$ is sufficiently close to $1$ 
(such that $t_a < \pi$). For the temperatures we plot, $t_a \sim 10^{-1}$ for the vector 
model, while it is far greater than $\pi$ in the adjoint model. Consequently, the adjoint 
correlators agree almost perfectly with the unconstrained behavior.
\end{description}

\section{Finite volume effects in the high temperature phase}\label{sec:non-classical}
In some regimes, we can probe physics that is hard to understand from a picture of a simple 
uniform source at the center of an asymptotically thermal AdS space. From our previous 
observation of an emergent length scale in spatial correlators 
\cite{AmadoSundborgThorlaciusWintergerst2017}, one expects a corresponding time scale,  
$t_a$, beyond which the deviation of the eigenvalue distribution from a $\delta$-function becomes visible. We have explicitly seen the emergence of such a time scale in our  numerical data of the previous section. There is a relatively simple conceptual argument for this scale, which can 
be made more quantitative for certain parameter values. The new length and time scales are 
proportional to the radius of the $d-1$-sphere and therefore only visible at finite spatial volume.

In the expression \eqref{eq:reallamapp} for the correlation function $G(t)$,
\begin{equation}
\int d\lambda\,\rho(\lambda)\left(\sum_{m=-\infty}^{\infty} \frac{e^{i m \lambda}}{\left(\sin^{2}
\left(\frac{t+i\beta m}{2}\right)\right)^{\frac{d-2}{2}}}\right)^2
= 
\int d\lambda\,\rho(\lambda)\left(\sum_{m=-\infty}^{\infty} 
\frac{e^{i m\beta \frac{\lambda}{\beta}}}{\left(\sin^{2}
\left(\frac{t+i\beta m}{2}\right)\right)^{\frac{d-2}{2}}}\right)^2 \,,
\end{equation}
the summation over $m$ is a Fourier series representation of a function 
$f_{t}^{(d)}(\lambda/\beta)$.
A $\delta$-function limit of the eigenvalue distribution corresponds to expanding 
$f_t$ in powers of $\lambda$, which can only yield a good approximation if 
$\lambda \ll \beta$ for all $\lambda$ in the support of $\rho(\lambda)$. 
Thus, in the absence of extra conditions on $t$,  $\lambda_m \ll \beta$ is required. 

The consequences of the above inequality vary with dimension. From our expressions 
\eqref{eq:lambdam} for the asymptotic values for $\lambda_m$, we conclude that the 
requirement is violated at all temperatures for the vector model in $d = 3$. 
For the vector model in $d=4$, it is fulfilled at temperatures that are exponentially 
large in $N$, while in higher even dimensions, it is easily satisfied high above the 
phase transition, measured on the scale of $T_{c}$. In the adjoint model on the 
other hand, the $\delta$-function approximation works well already at temperatures 
only slightly above the transition. See Table~\ref{tab:tatab} for a summary of the relevant temperature scales in different dimensions.

A more quantitative analysis can be carried out. Choosing for 
illustrative purposes $d=4$, we consider the expression in real $\lambda$-space 
Eq.~\eqref{eq:reallampoisseval} (here for the vector model), slightly reformulated to read
 \begin{equation}
 \label{eq:gtaddscale}
 G(t)  = \frac{4\pi^2}{\beta^4}
 \int d\lambda\,\rho(\lambda)\, e^{-\frac{2\lambda  t}{\beta }}\left[\lambda 
 \left(\coth\left(\frac{\pi t}{\beta}\right)+\coth\left(\frac{\pi\lambda}{\beta}\right)\right)
 +\pi\text{csch}^2\left(\frac{\pi t}{\beta}\right)\right]^2  \,.
 \end{equation}
By expanding $\coth\left(\frac{\pi\lambda}{\beta}\right)$ in a series of exponentials, 
rewriting powers of $\lambda$ as time derivatives of $e^{-\lambda t/\beta}$, and using 
$\rho(\lambda)|_{|\lambda| > \lambda_m} = 0$, we observe that all terms in the above 
expression are of the form
\begin{equation}
f(t)\, \partial_t^n \left(\int_{0}^\infty d\lambda\,\rho(\lambda) 
\left(e^{-\frac{2\lambda}{\beta}(t+\pi k)}\pm e^{-\frac{2\lambda}{\beta}(-t+\pi k)}\right)\right)\,,
\end{equation}
with different functions $f(t)$ and non-negative integers $n$ and $k$. Recognizing the 
Laplace transform of $\rho$, denoted by $\bar\rho$, this corresponds to
\begin{equation}
\label{eq:lapgenform}
f(t)\, \partial_t^n \left(\bar\rho\left(\frac{2(t+\pi k)}{\beta}\right)
\pm \bar\rho\left(\frac{2(-t+\pi k)}{\beta}\right)\right)\,.
\end{equation}
The emergence of the additional time scale $t_a$ is evident already at this stage, since 
$\bar\rho$ inevitably depends on the scale $\lambda_m$. We can make this statement 
more precise, however.  At large temperatures, under the condition\footnote{This is 
precisely the case when the crossover time $t_a$ is below the light crossing time of the 
sphere.} that $\lambda_m \gg \beta$, the leading contribution to $G(t)$ comes from 
those terms in \eqref{eq:lapgenform} for which $k = 0$. There, an explicit integration of 
Eq.~\eqref{eq:gtaddscale}, obtained by writing it as an integral over positive $\lambda$ 
only and setting $\coth\left(\frac{\pi\lambda}{\beta}\right) = 1$, yields
\begin{equation}
\label{eq:glapsimp}
G(t) \approx \frac{4\pi^4}{\beta^4}\sum_{n=0}^2 \pi^{-n} {2 \choose n} 
\text{csch}^{4-n}\left(\frac{\pi t}{\beta}\right)\left(e^{-\frac{n\pi t}{\beta}}\bar\rho^{(n)}
\left(-\frac{2 t}{\beta}\right)+(-1)^n e^{\frac{n\pi t}{\beta}}\bar\rho^{(n)}
\left(\frac{2 t}{\beta}\right)\right)
\end{equation}
We can infer almost all essential details of the correlator from here, using the following 
generic features of the Laplace transform of the eigenvalue density:
\begin{enumerate}[label=(\roman*)]
	\item \label{it:small} $\bar \rho^{(n)}(x) \sim (-\lambda_m)^n$ for $|x| \ll \lambda_m^{-1}$, 
	\item \label{it:large} $\bar \rho(x) \to (\lambda_m x)^{-1}+ \frac{e^{-x \lambda_m}}{x^2 
	\lambda_m^2}$ for $|x| \gg \lambda_m^{-1}$.
\end{enumerate}
These properties follow directly from the known behavior of the eigenvalue distribution 
in real $\lambda$-space. For $x \ll \lambda_m^{-1}$, their Laplace transforms follow that 
of a $\delta$-function, with corrections parametrized by the only scale in the problem, 
$\lambda_m$. The large-$x$ fall-off, on the other hand, follows from the fact that the 
distributions are approximately constant for sufficiently small $\lambda$. It is here that 
the deviation from a $\delta$-function distribution becomes truly significant. 
The exponential contribution follows from the cut-off of $\rho(\lambda)$ at 
$\lambda = \lambda_m$. Note that the exponential growth for $x<0$ will be 
overpowered by the exponentially decaying factors in Eq.~\eqref{eq:glapsimp}, 
as we see below. 

Using the above, we obtain the following behavior for the correlation function.
At very short times, the vacuum contribution dominates, hidden in the $n=0$-term of the 
above sum. Once $t$ becomes of order of $\beta$, there is a period of exponential decay, 
easily seen by inserting case~\ref{it:small} into Eq.~\eqref{eq:glapsimp}. 
The dominant contributions are
\begin{equation}
\label{eq:exponential_d4}
G(t) \approx \frac{16\pi^2}{\beta^4}\left(4\pi^2 e^{-\frac{4\pi t}{\beta}} 
+ 4\pi\lambda_m e^{-\frac{2\pi t}{\beta}} + \lambda_m^2\right)\,.
\end{equation}
As we see, the exponential decay is eventually stopped by the constant term, 
at $t \sim \log{\lambda_{m}}/T \sim \log{(T/T_{c})}/T$ for $T > T_{c}$, where there 
is a gap in the eigenvalue density. This constant in turn decays once the central part 
of the eigenvalue distribution becomes accessible, at $t/\beta \sim \lambda_m^{-1}$ or 
equivalently $t \sim t_{a}$ with 
\begin{equation}
t_{a}\equiv \frac{\beta}{\lambda_m} \,. \label{eq:crossover_time}
\end{equation}
We obtain the late time behavior by inserting case~\ref{it:large} into Eq.~\eqref{eq:glapsimp}. 
All exponential pieces are negligible, in particular since $t > t_a \gg \beta$ and 
$\lambda_m \leq \pi$. The only remaining contribution gives
\begin{equation}
\label{eq:ren_power}
G(t) \approx \frac{4\pi^2}{t^3\beta\lambda_m}\,.
\end{equation}
Comparison with Eq.~\eqref{eq:lateLowTpower} for $d=4$ shows the same decay 
as for the low temperature correlator, with a temperature dependent renormalization factor. 
We have thus analytically reproduced all features of Fig.~\ref{fig:decay}.

As promised earlier, Eq.~\eqref{eq:glapsimp} also holds the key to seeing that the low 
temperature behavior of two successive power law regimes, as described in section~\ref{sec:low}, holds up until the 
phase transition. The low temperature phase is characterized by the absence of the small parameter $\lambda_m$ and the  Laplace 
transform of $\rho$ is always described by $\bar\rho(x) \sim x^{-1}$. This implies that the $n=2$ term in Eq.~\eqref{eq:glapsimp} always dominates the thermal contributions, since derivatives of $\bar\rho$ are not suppressed. There is thus no leading exponential decay and the late time fall-off is given by the power law that is encoded in the Laplace transform, according to Eq.~\eqref{eq:ren_power}.
Conversely, the argument confirms that the appearance of a dominant exponential 
regime is directly linked to the eigenvalue distribution becoming zero on a finite interval.

Similar arguments as above can be made for $d=3$, although they are somewhat less 
clear due to the lack of description of the correlator in terms of elementary functions. 
A proof of concept goes as follows. For small $\beta$, the correlator is approximately 
described by Eq.~\eqref{eq:oddlamriem}, which after integration reads
\begin{equation}
\label{eq:oddlamriemint}
G(t)  = -\frac{1}{2}
\int d\lambda\,\rho(\lambda)\left(\frac{1}{\sin\left(\frac{t}{2}\right)}-\frac{2i}{\beta} 
e^{-\frac{t\lambda}{\beta}}\,\text{Re}B_{e^{i t}}\left(\frac{1}{2}
+\frac{i\lambda}{\beta},0\right)\right)^2\,.
\end{equation}
For $t \ll 1$, the dominant contribution to the above comes from the $\lambda \to 0$ 
limit of the Beta function, since the latter can be expanded in a power series in $\lambda$, 
which converges quickly for $t \ll 1$. Higher powers of $\lambda$ are then suppressed by 
powers of $\lambda_m$ in the integrated expression. Moreover, there are cancellations 
which suppress the contributions from the integration over negative $\lambda$.
In the small $t$ limit we obtain
\begin{align}
G(t)  &\approx -2
\int d\lambda\,\rho(\lambda)\left(\frac{1}{t}+\frac{2i}{\beta} 
e^{-\frac{t\lambda}{\beta}}\log\left(\frac{t}{4}\right)\right)^2\nonumber\\
&\approx-2\left(\frac{1}{t^2} + \frac{4i}{t\beta} \log\left(\frac{t}{4}\right)
\bar\rho\left(\frac{t}{\beta}\right)-\frac{4}{\beta^2} \log^2\left(\frac{t}{4}\right)
\bar\rho\left(2\frac{t}{\beta}\right)\right)
\label{eq:oddlamsmallt}\,.
\end{align}
We thus see that the logarithmic decay is accelerated once $t \gtrsim t_a$, 
in agreement with Figs.~\ref{fig:fun3dreal} and \ref{fig:fun3dret}.

\renewcommand{\arraystretch}{1.5}
\begin{table}
	\centering
	\begin{tabular}{cccc}
		& $d = 3$ & $d = 4$ & $d \geq 5$ \\ \hline
		$T_c$	& $N^{\frac{1}{2}}$ & $N^\frac{1}{3}$ & $N^\frac{1}{d-1}$ \\ 
		$t_{a}$	& $N^{-\frac{1}{2}}$	&$N^{-\frac{1}{3}}\log^{1/3}\left(\frac{T}{T_{c}}\right)$&				$N^{-\frac{1}{d-1}}\left(\frac{T}{T_{c}}\right)^{\frac{d-4}{3}}$\\
		$T_a$	& ---  & $N^\frac{1}{3} e^N$ & $N^\frac{1}{d-4}$ \\  \hline
	\end{tabular}
	\caption{Scaling of the critical temperature $T_c$, crossover time $t_{a}$ and the 
		constraining temperature $T_a$ in the vector model for dimensions $d \geq 3$. 
		The critical temperature follows directly from Eq.~\eqref{Tcritical}, while $T_a$ is 
		obtained from the condition $t_a \sim 1$, inserting the asymptotic expressions for 
		$\lambda_m$, Eq.~\eqref{eq:lambdam} in Eq.~\eqref{eq:crossover_time}.	\label{tab:tatab}}
\end{table}

To summarize, the emergence of an additional time scale follows in the same way as an 
additional length scale in \cite{AmadoSundborgThorlaciusWintergerst2017} from the fact 
that the integrand in all expressions for the correlator is a function of $\lambda/\beta$, and 
may be Taylor expanded in $\lambda$ for general $t$ only if $\lambda_m/\beta = t_a^{-1} \ll 1$. 
We have moreover demonstrated analytically that the correlators become sensitive to the core 
of the eigenvalue distribution at times $t > t_a$.
This leaves visible imprints on the eigenvalue distributions if $t_a$ is smaller than the 
light crossing time of the sphere. The temperature dependence of $\lambda_m$ implies 
that for $d > 3$, there is only a finite range of temperatures, $T_c \leq T \leq T_a$, in which 
this is true. We summarize the dependence of $T_c$ and $T_a$ on $N$ and on $d$ for 
the vector model in Table \ref{tab:tatab}. 

We see a manifestation of the scale $t_a$ for a parametrically large window of temperatures 
which extends over all $T \geq T_c$ in $d=3$, to exponentially high temperatures in $d = 4$ 
and to temperatures that grow like a positive power of $N$ in $d \geq 5$.
For adjoint scalars, on the other hand, since $T_c = {\cal O}(1)$, $T_a \simeq T_c$ for 
all $d$.

\section{Discussion}\label{sec:discussion}
In this section we summarize our findings, discuss connections to
other developments and indicate some directions for future work.
A concrete goal of our work is to assess the status of black holes in 
higher spin bulk theories that are in effect defined by the free gauge 
theories considered here. 
All our calculations have been carried out on the field theory
side of the gauge/gravity duality and we work in the limit of vanishing 't~Hooft 
coupling where the bulk interpretation of results is not {\it a priori} clear.
We adopt an operational strategy and look for 
signatures in boundary correlation functions that are known, from 
previous studies at large 't~Hooft coupling, to correspond to specific 
bulk features, e.g.\ black holes. In other words, if a bulk object looks 
like a black hole from the boundary perspective we call it black hole-like. 

We work at finite temperature and finite volume throughout and 
all our results are obtained in a large~$N$ limit, where the physics can be
analyzed in terms of saddle point solutions for the density of eigenvalues 
$\rho(\lambda)$ of the gauge holonomy around the thermal circle. The 
eigenvalue density determines the large~$N$ phase diagram of the theory 
and also governs the behavior of correlation functions of gauge invariant
operators. The critical temperature for phase transitions on the unit 
$d-1$-sphere is $T_{c} \sim 1$ for a singlet model with scalar matter in 
the adjoint representation and $T_{c} \sim N^{1/(d-1)}$ for a model 
with vector representation matter. In terms of bulk physics these critical 
temperatures correspond to $T_{c} \sim R_{AdS}^{-1}$ and $T_{c} \sim m_{Pl}$, 
respectively.

We have found evidence for black hole-like behavior previously observed in 
gauge/gravity duality in the limit where the bulk side of the duality is well described
by classical gravity. It is useful to spell out a large $N$ temperature ladder for 
discussing these results.
\begin{itemize}
\item[]
{\it Low temperature}: $T < T_c$ in adjoint model, $T\leq O(1)$ in vector model. 
Uniform eigenvalue density $\rho(\lambda)=\frac{1}{2\pi}$
in both models. 
\item[]
{\it Super-AdS temperature}: $1\ll T\leq O(T_c)$ in vector model. The eigenvalue density is no
longer uniform but no gap has opened up. This range is absent in the adjoint model where
$T_c\sim 1$.
\item[]
{\it High temperature}: $T_c<T$ in both vector and adjoint models. The eigenvalue density
has developed a finite gap where $\rho(\lambda)=0$.
\item[]
{\it Ultra-high temperature}: $T_{c}< T_a\ll T$ in both types of models, 
where $T_{a}$ is given in Table~\ref{tab:tatab}. It is defined to ensure that the eigenvalue density at higher temperatures  is
well approximated by a Dirac delta function (not strictly attainable in $d=3$). 
\end{itemize}
Our main results can be summarized as follows in relation to these temperature scales:
\begin{itemize}
\item 
We find evidence for evanescent modes 
(section \ref{sec:evanescence}) of the type pointed out in \cite{ReyRosenhaus2014}. 
Evanescent behavior occurs in any $d \geq 3$ at super-AdS, high and ultra-high 
temperatures in the vector model and at all temperatures above the critical temperature 
in the adjoint model.
\item We see a period of exponential decay in time of correlators (section \ref{sec:exp_decay}) of the 
type pointed out in \cite{Maldacena2003} for two boundary dimensions. We  demonstrate 
this behavior explicitly for any even $d$ at all temperatures above the phase transition in both vector and adjoint models. It is also visible in our
numerical results for $t \lesssim \log{(T/T_{c})}/T$ provided $T > T_{c}$. 
\end{itemize}
The exponential decay is not immediately visible and it does not continue indefinitely. 
At very early times, short compared to the inverse temperature, the two-point 
correlation function decays according to a power law governed by the underlying scale 
symmetry of the model. Exponential decay sets in at $t\sim \beta$ and the correlation
function settles to a constant value at $t\sim (\log T)/T$, which is maintained for a while. 
Eventually the periodicity of free wave propagation on the sphere sets in and 
the correlator returns to its initial value at $t=2\pi$ in our units (see Figure~\ref{fig:fun3dreal}).

We note that in even $d$, evanescent modes are evident at lower temperature than 
exponential time decay in vector models but appear together with exponential time decay 
at the critical temperature in adjoint models. 
In odd $d$ we still see evanescent behavior but exponential decay in time is absent. 
Instead, we find what may be a new signal of black hole behavior in odd boundary 
dimensions:
\begin{itemize}
\item  The boundary Green's function has support inside the light cone\footnote{As 
opposed to only on the light cone.} in odd $d$ (section \ref{sec:retarded}). This occurs at 
super-AdS, high, and ultra-high temperatures in vector models and at $T>T_c$ in  
adjoint models.
\end{itemize}
Finally, we have found a new time scale at which the time-dependence of boundary 
correlation functions changes qualitatively. This behavior is only detectable at high 
temperatures and finite 
volume, {\it i.e.} away from the planar limit in the bulk:
\begin{itemize}
\item For times beyond a new time scale $t_{a}$, boundary correlators return to their 
short-time, thermal AdS form, up to a temperature dependent normalization factor 
(see section \ref{sec:non-classical}, especially Table \ref{tab:tatab}). 
This behavior is seen at all high temperatures in $d=3$, but only over a finite range of 
temperatures, $T_{c}<T<T_{a}$, in $d\geq 4$.
\end{itemize}
In the $d=3$ vector model, we find the time scale $t_a = T_{c}^{-1}\sim N^{-1/2}$, 
and more generally, $t_{a}$ is given in Table \ref{tab:tatab}.

By the usual AdS/CFT dictionary, long boundary distances or late times translate to 
bulk features that are far removed from the boundary. Accordingly, we expect long-time 
correlators to probe more deeply into the bulk, while short-time correlators are mostly 
sensitive to propagation close to the boundary. Thus, very short $t\ll\beta$ only sense 
asymptotic AdS propagation, intermediate times $\beta\ll t \ll t_a$ can disclose the 
presence of black hole-like objects, while later times $t \gg t_a$ probe deep bulk 
features which are not accessible in a planar geometry.

Earlier studies ({\it e.g.} \cite{ShenkerYin2011}) have cautiously refrained from interpreting the 
vector model phase transition as a sign of black holes 
in higher spin gravity. Our findings crucially add spatial extent to the picture.
Our previous work \cite{AmadoSundborgThorlaciusWintergerst2017} revealed a boundary 
spatial scale where equal time correlation functions at high temperature begin to 
depart from thermal AdS behavior, and we have now also confirmed a 
corresponding time scale consistent with bulk objects that grow with temperature. 
Since this boundary scale only comes into play at high temperature 
$T > T_{c}\sim m_{Pl}$, the corresponding black hole-like objects in the bulk have a minimum size, 
which scales with $N$ in the same way as the Planck mass. Thus, the objects appear to 
become larger with increasing temperature and with decreasing gravitational coupling.
The enormous size of these higher spin black hole-like objects demystifies their Planck scale 
temperature. They are analogous to standard large AdS black holes in Einstein gravity, 
whose Hawking temperature grows without bound with increasing black hole size.
For a large enough AdS black hole, the Hawking temperature becomes trans-Planckian but
this is not problematic as this is not the temperature measured by observers in free fall in the 
bulk spacetime geometry \cite{HemmingThorlacius2007,BrynjolfssonThorlacius2008}.
Similar considerations presumably apply to very large higher spin black holes.

In free singlet theories, periods of exponential decay of correlations of 
approximately local field operators find their origin in an intricate cancellation of phases.
In case of a two-dimensional boundary, infinite dimensional conformal symmetry on the cylinder 
dictates \cite{Cardy1984,BirminghamSachsSolodukhin2002} that thermal correlation functions in 
free boundary theories exhibit similar exponential fall-off behavior as their strongly interacting 
counterparts \cite{Maldacena2003}, even though they do not thermalize in the usual sense. 
On the torus, corresponding to finite size, the argument indicates exponentially decaying terms 
in the high temperature limit.
Such arguments are not applicable in higher dimensions, and there is no a priori reason for an exponential fall-off. Moreover, delocalized Fourier space operators exhibit power law tails (\cite{HartnollKumar2005} and Appendix \ref{app:fourier}), much like the low temperature autocorrelators we discussed in section 
	\ref{sec:low}. The fact that thermal Green's functions of quasi-local operators nevertheless display periods of exponential decay may be taken as a sign that bulk higher spin theories retain features of their gravitational ancestors. In a sense, nontrivial phase cancellations are the only way for a 
	free boundary theory to produce behavior similar to
	a strongly coupled CFT with a gravitational dual.

Since our models are explicitly unitary and propagation in classical black hole backgrounds 
is known to break unitarity,  our results must deviate from classical black hole results in a 
way that restores unitarity. In even dimensions we indeed found a cross-over from exponential 
decay to a constant correlator at a time $t \sim \log T/T$ 
valid above the phase transition. Qualitatively, the crossover is reminiscent of the results 
in \cite{Maldacena2003} where an exponential decay crosses over to a non-decaying regime 
at late time, explained by the contribution from an alternative bulk manifold filling the same 
boundary. The sharpness of the transition visible in Fig.~\ref{fig:decay} suggests such an 
interpretation, but in the present higher spin case, we have no supporting bulk calculation. 
In vector models there is a later time $t_a$ when the singlet constraint comes into operation
and the correlation function develops another power law fall-off.  This time the power law 
exponent is the same as in the low-temperature phase, but the correlation function picks up a 
temperature dependent normalization factor. 
In odd dimensions, a case not detailed in \cite{Maldacena2003}, we see a related crossover 
between two different power laws. We note that at high temperature this crossover coincides
with a zero of the real part of the correlator (see Fig.~\ref{fig:fun3dreal}).

In two boundary dimensions the advanced machinery of $d=2$ conformal field theory 
has been applied to the tricky problem of reconciling the unitarity of CFT with the known, 
but approximate information loss in bulk black hole physics 
\cite{FitzpatrickKaplanLiWang2016,DyerGur-Ari2017,CampoMolina-VilaplanaSonner2017,ChenHussongKaplanLi2017}. 
We have glimpsed related behavior in higher dimensions for very particular models. 
It would be interesting to study these issues in more general high-dimensional models
and look for similar effects. 

The intriguing return to thermal AdS-like correlation functions
after a period of black hole-like correlator behavior
also deserves further study. 
We have observed this in the high temperature phase both at late 
times and large angle in low dimensions, suggesting that we are probing deep into the bulk. 
We have established that the high temperature phase displays black hole-like features, like
evanescent modes, and a closer study of deep probes might reveal the inner workings 
of emergent black holes in higher spin gravity. 

\begin{acknowledgments}
This work was supported in part by the Swedish Research Council through the 
Oskar Klein Centre, under 
contract 621-2014-5838 and under contract 335-2014-7424, the Icelandic Research Fund grant 163422-052 and
the University of Iceland Research Fund.
\end{acknowledgments}

\appendix
\section{Saddle point evaluation of path integrals and eigenvalue distributions}
\label{sec:ev_dists}

We wish to evaluate the Euclidean path integral of a $U(N)$ gauge theory on 
$S^{d-1}\times S^1$ with scalar matter fields in the adjoint or fundamental 
representation,
\begin{equation}
\label{zpathintegral}
	Z[\beta] = \int {\cal D}A_\mu {\cal D}\phi {\cal D}\phi^\dagger 
	e^{-S[A_\mu, \phi, \phi^\dagger; \beta]}\,,
\end{equation}
in the limit of vanishing gauge coupling. For a detailed derivation, including
gauge fixing and evaluating the corresponding Faddeev-Popov determinants, 
see {\it e.g.}~\cite{AharonyMarsanoMinwallaPapadodimasVan-Raamsdonk2004}. 
We do not repeat the derivation here but merely outline the main steps in order 
to motivate the eigenvalue distributions that we work with in the rest of the paper.

On a compact spatial manifold 
gauge invariance imposes the vanishing of all gauge charges and the 
only physical states are gauge singlets. In the free theory limit the gauge
field decouples from the matter fields apart from a zero mode,
\begin{equation}
\alpha=\frac{1}{Vol(S^{d-1})}\int_{S^{d-1}} A_t,
\end{equation}
that enters into gauge covariant derivatives, $D_{t} = \partial_t + i\alpha$ in the 
vector model and $D_{t} = \partial_t + i[\alpha,\cdot]$ for an adjoint 
scalar~\cite{AharonyMarsanoMinwallaPapadodimasVan-Raamsdonk2004}. 
The zero mode also gives rise to a gauge holonomy, $U=e^{i\beta \alpha}$, 
going around the thermal circle. 
The path integral over the gauge
field reduces to a $U(N)$ matrix integral over the zero mode, which in
turn reduces to an integral over the eigenvalues $e^{i\lambda_i}$ of the 
gauge holonomy matrix,
\begin{equation}
\label{eq:genfun}
Z[\beta] = \frac{1}{N!} \int \left(\prod_i d\lambda_i\right) 
\exp\Bigg[  \sum_{i\neq j} \ln\left|\sin\left(\frac{\lambda_i - \lambda_j}{2}\right)\right|
- \sum_i S_s[\lambda_i] \Bigg] \,.
\end{equation}
The matter contribution to the action $S_s[\lambda]$ comes from integrating
out the scalar field in the path integral \eqref{zpathintegral}, keeping track of the gauge 
field zero mode in the covariant derivative. In the vector model it is
given by
\begin{equation}
S_s[\lambda_i] = -2N_f \sum_{k=1}^{\infty}\frac{1}{k}z_S^d(x^k) \cos(k\lambda_i) \,,
\end{equation}
and in the adjoint model by
\begin{equation}
S_s[\lambda_i] =  -\sum_{k=1}^{\infty} \frac{1}{k} z_S^d(x^k) \sum_ j \cos(k(\lambda_i 
- \lambda_j))\,,
\end{equation}
where $z_S^d(x)$ is the scalar one-particle partition sum on a $d-1$-dimensional sphere,
\begin{equation}
z_S^d(x)=x^{\frac{d}{2} - 1}\frac{1+x}{(1-x)^{d-1}}\,,
\end{equation}
with $x = e^{-\beta}$.

In the large $N$ limit the integral over eigenvalues can be evaluated in a 
saddle point approximation. For this it is convenient to work with the 
eigenvalue distribution $\rho(\lambda)=\frac{1}{N} \sum_{i=1}^N \delta(\lambda-\lambda_i)$
or equivalently its Fourier cosine coefficients,
\begin{equation}
\rho_k =\int_{-\pi}^\pi d\lambda\,\rho(\lambda) \cos{k\lambda}
=   \frac{1}{N} \sum_{i=1}^N \cos{k\lambda_i} \,.
\end{equation}
In the large~$N$ limit the eigenvalue density approaches a continuum. It is 
positive semi-definite, 
\begin{equation}
\label{poscond}
\rho(\lambda) \geq 0 \,,
\end{equation}
and normalized to one in our conventions,
\begin{equation}
\label{normalcond}
\int_{-\pi}^\pi d\lambda\,\rho(\lambda) = 1\,.
\end{equation}
By using the relation 
\begin{equation}
\ln\left|\sin\left(\frac{\lambda_i - \lambda_j}{2}\right)\right| =  -\log 2 - \sum_{k=1}^\infty \frac{1}{k}\cos(k(\lambda_i - \lambda_j))\,,
\end{equation}
and assuming that the eigenvalue distribution is symmetric under sign reversal
$\lambda\rightarrow -\lambda$, 
the integrand in \eqref{eq:genfun} can be re-expressed in terms of an effective action 
involving the eigenvalue distribution.\footnote{Assuming a symmetric eigenvalue distribution
leads to the apparent breaking of the $\mathbb{Z}_n$ center symmetry of the adjoint model.
The symmetry can be restored by summing over $\mathbb{Z}_n$ images of the saddle point
but this only leads to an overall factor and does not affect our results. In the vector model
the center symmetry is broken from the start by the matter sector.}

In the vector model one finds
\begin{equation}
\label{eq:act_fun}
S = N^2\sum_{k=1}^\infty\frac{\rho_k}{k}\left(\rho_k - 2\frac{N_f}{N}z_S^d(x^k)\right)\,.
\end{equation}
which is minimized by 
\begin{equation}
\rho_k = \frac{N_f}{N}z_S^d(x^k)\,.
\end{equation}
The corresponding eigenvalue density, given by 
\begin{equation}
	\rho(\lambda) = \frac{1}{2\pi} + \frac{N_f}{N}
	\sum_{k=1}^{\infty} z_S^d(x^k) \frac{1}{\pi}\cos(k\lambda)\,,
\end{equation}
appears in equation \eqref{eq:dist_sum} in the main text. The constant term in 
the eigenvalue density is needed for the normalization condition \eqref{normalcond}.
As discussed in Section~\ref{sec:phase_transition}, the positivity condition \eqref{poscond} 
is satisfied for all $\lambda\in[-\pi,\pi]$ at low temperatures but there is a phase transition
and in the high temperature phase the eigenvalue distribution develops a gap, {\it i.e.} there 
is a finite range $\lambda_m\leq|\lambda|\leq\pi$ where $\rho(\lambda)=0$. 
The critical temperature \eqref{Tcritical} is determined by the condition $\rho(\pm\pi)=0$.

At high $T$, but still below $T_c$, the eigenvalue density can be expressed in terms 
of a polylogarithm \cite{AmadoSundborgThorlaciusWintergerst2017},
\begin{equation}
	\label{eq:dist_polylog}
	\rho(\lambda) = \frac{1}{2\pi} +
	A_d\left(\frac{T}{T_c}\right)^{d-1}
	\text{Re}\left(\text{Li}_{d-1}(e^{i\lambda})\right),
\end{equation}
where $A_d= \frac{1}{2\pi\left(1-2^{2-d}\right)\, \zeta(d-1)}$.
At high temperatures beyond the phase transition the eigenvalue density is still given by
a polylogarithm but now it vanishes over a
finite interval,
\begin{equation}
\rho(\lambda) = 
\begin{cases}
A_d\left(\frac{T}{T_c}\right)^{d-1}\left(\text{Re}\, \text{Li}_{d-1}(e^{i\lambda}) 
- \text{Re}\, \text{Li}_{d-1}(e^{i\lambda_m})\right)&\text{for } |\lambda| \leq \lambda_m\,,\text{and}\\
0&\text{otherwise}\,.
\end{cases}
\end{equation}
The maximal eigenvalue $\lambda_m$ is fixed by the normalisation condition, 
$\int_{-\lambda_m}^{\lambda_m}\rho(\lambda)d\lambda=1$, which can be expressed 
as follows in terms of polylogarithms,
\begin{equation}
\label{eq:norm_d}
\text{Im}\, \text{Li}_{d}(e^{i\lambda_m})-\lambda_m\text{Re}\, \text{Li}_{d-1}(e^{i\lambda_m})
=\frac{1}{2A_d} \left(\frac{T_c}{T}\right)^{d-1}\,.
\end{equation}

In the limit of very high temperature the eigenvalue distribution is narrowly peaked around 
the origin. One obtains the scaling behavior \eqref{eq:lambdam} for the width of 
the eigenvalue distribution at $T\gg T_c$,
and the eigenvalue distribution itself can be approximated by 
\begin{equation}\label{eq:highTrho}
\rho(\lambda) = \frac{A_d}{2} \left(\frac{T}{T_c}\right)^{d-1}
\begin{cases}
\pi(\lambda_m - |\lambda|)&\text{for } d = 3\,,\\
\frac{1}{2}\left(\lambda_m^2(3-2\log\lambda_m)
-\lambda^2(3-2\log|\lambda|)\right)&\text{for } d = 4\,,\\
\zeta(d-3) (\lambda_m^2 - \lambda^2)&\text{for } d \geq 5\,.
\end{cases}
\end{equation}

In the adjoint model the effective action takes the form
\begin{equation}
\label{eq:act_adj}
S = N^2 \sum_{k=1}^\infty\frac{(1 - z_S^d(x^k))}{k} \rho_k^2\,.
\end{equation}
As long as $z_S^d(x^k) < 1$, the favored configuration is a homogeneous 
distribution $\rho(\lambda)=\frac{1}{2\pi}$, which has $\rho_k=0$ for all $k\geq 1$.
There is a critical temperature at which $z_S^d(x_c) = 1$ and the first non-trivial 
Fourier cosine mode of the eigenvalue distribution condenses.
The precise value of the critical temperature depends on the number of spatial
dimensions but $T_c\sim \mathcal{O}(1)$ for all $d$. 
As we go to higher temperatures, above the phase transition, the higher modes 
successively condense and the detailed implementation of the positivity constraint 
on $\rho(\lambda)$ becomes somewhat intricate. The approximate expression 
\eqref{approxrho} for the eigenvalue density, developed 
in~\cite{AharonyMarsanoMinwallaPapadodimasVan-Raamsdonk2004}, simplifies the 
analysis considerably and is sufficient for our purposes in this paper. In particular, it
allows one to infer the correct scaling behavior of the width of the eigenvalue 
distribution at high temperatures, $\lambda_m\rightarrow T^\frac{1-d}{2}$. 

\section{Time dependent Green functions}
\label{app:greenfcn}

In singlet models, the main subtlety in constructing time dependent correlators 
lies in the implementation of the constraint. In the formalism of Schwinger and 
Keldysh \cite{Schwinger1961,Keldysh1964}, this is achieved through various 
insertions of $\delta$-functions that project onto physical states. A careful treatment 
of the integration along the complex time contour then gives rise to a generating 
functional of time-dependent correlation functions that obey the singlet conditions.

The free singlet models considered in the present work offer a significant 
shortcut via straightforward analytic continuation of the Euclidean propagator to the real time 
domain. In the vector theory the Euclidean 
propagator takes the form~\cite{AmadoSundborgThorlaciusWintergerst2017},
\begin{equation}
G^{AB}_{lM}(n) =
\sum_{j=1}^N \frac{\Psi_{j}^A (\Psi^{B}_j)^\dagger }
{\beta^{-2}\left(2\pi n + \lambda_j\right)^2 + E_\ell^2}\,,
\label{eq:gf_vec}
\end{equation}
in frequency/angular momentum space. Here, $E_\ell = (\ell+(d-2)/2)$ and 
$\Psi_{i}$ is an eigenvector of the gauge holonomy matrix that belongs 
to the eigenvalue $e^{i\lambda_i}$. The eigenvectors are taken to be normalized, 
\begin{equation}
\label{eq:evec_norm}
\sum_{A=1}^N\Psi_{i}^A (\Psi_{j}^A)^\dagger = \delta_{ij}\,,\hspace{3em}
\sum_{i=1}^N\Psi_{i}^A (\Psi_{i}^B)^\dagger = \delta^{AB}\,.
\end{equation}

Analytic continuation, 
$\beta^{-2}\left(2\pi n + \lambda_i\right)^2 \to -(\omega + i\epsilon)^2$ yields a
momentum space expression for the retarded propagator,
\begin{equation}
G^{AB}_{R,lM}(\omega) = \frac{\delta^{AB}}
{-(\omega + i\epsilon)^2 + E_\ell^2}\,.
\label{eq:gf_r_vec}
\end{equation}
Using well-known thermal field theory relations between the time ordered 
propagator, the retarded propagator, and its discontinuity in presence of an imaginary 
chemical potential,\footnote{A nice derivation can for 
instance be found in \cite{LaineVuorinen2016}.}
\begin{equation}
G_T^{AB} = -i\,G_R^{AB} - 2\,\text{Im}\,G_R^{AC}\sum_{j=1}^{N} n_B \left(\omega
+i \frac{\lambda_j}{\beta}\right)\Psi_{j}^C (\Psi^{B}_j)^\dagger ,
\end{equation}
one obtains after some manipulations the following time ordered propagator,
\begin{equation}
G^{AB}_{T,lM}(\omega) = \frac{i}{\omega^2 - E_\ell^2 + i\epsilon}\delta^{AB} 
+ 2\pi\delta(\omega^2 - E_\ell^2)\,\sum_{j=1}^{N}n_B\left(|\omega|
+i \frac{\omega}{|\omega|}\frac{\lambda_j}{\beta}\right)\Psi_{j}^A (\Psi^{B}_j)^\dagger\,.
\label{eq:gf_t_vec}
\end{equation}
The fact that the eigenvalues of 
the holonomy matrix appear as an imaginary chemical potential is a well known feature 
of thermal correlation functions in gauge theories and follows from the  
holonomy matrix being the Lagrange multiplier that implements the gauge singlet 
constraint \cite{HartnollKumar2005}.
Fourier transforming Eq.~\eqref{eq:gf_t_vec} in time yields the mixed space expression,
\begin{multline}
G^{AB}_{T}(t,E) = \frac{1}{i} \frac{1}{2 E} 
\sum_{j=1}^N \Psi^{j}_A (\Psi^{j}_B)^\dagger 
\\
\times\left[e^{-i E t} \left(\Theta(t)+ n_B(E - i\frac{\lambda_j}{\beta})\right) + e^{i E t} \left(\theta(-t)+ n_B(E + i\frac{\lambda_j}{\beta})\right)\right]\,.
\label{eq:gf_t_vec_k}
\end{multline}
Finally, Eq.~\eqref{eq:gf_t_vec_k} can be transformed to yield the real space expression
(using time translational and rotational invariance to set one of the insertion points to $x = 0$)
\begin{multline}
G^{AB}_{T}(t,{\bf y}) = \frac{1}{i} \sum_{l,M} \frac{1}{2 E_l} Y_{l,M}({\bf y}) Y_{l,M}(0)
\sum_{j=1}^N \Psi^{j}_A (\Psi^{j}_B)^\dagger 
\\
\times\left[e^{-i E_l t} \left(\Theta(t)+ n_B(E_l - i\frac{\lambda_j}{\beta})\right) + e^{i E_l t} \left(\theta(-t)+ n_B(E_l + i\frac{\lambda_j}{\beta})\right)\right]\,.
\label{eq:gf_sphere}
\end{multline}

We can evaluate this expression further by using the fact that $Y_{l,M}(0) \propto \delta_{M,0}$ and inserting the form of the spherical harmonics for $M = 0$. 
First taking $t>0$ and writing the distribution functions as sums over exponentials 
\begin{multline}
G^{AB}_{T}(t,{\bf y}) = \frac{\Gamma(\sigma)}{4\pi^{\sigma+1}i} \sum_{l} C_{l}^{(\sigma)}(\cos \theta) 
\sum_{j=1}^N \Psi^{j}_A (\Psi^{j}_B)^\dagger 
\\
\times\left[e^{-i t(l+\sigma)} \left(1+ \sum_{n=1}^{\infty} e^{-\beta n(l+\sigma)+i n\lambda_j}\right) + e^{i t(l+\sigma)} \left( \sum_{n=1}^{\infty} e^{-\beta n(l+\sigma)-i n\lambda_j})\right)\right]
\label{eq:gf_sphere2.1}\,,
\end{multline}
we find 
\begin{multline}
G^{AB}_{T}(t,{\bf y}) = \frac{\Gamma(\sigma)}{4\pi^{\sigma+1}i}
\sum_{j=1}^N \Psi^{j}_A (\Psi^{j}_B)^\dagger 
\\
\times\left[
\frac{1}{(2\cos{t}-2\cos{\theta})^{\sigma}}+ 
\sum_{n=1}^{\infty} \frac{e^{i n\lambda_j}}{(2\cos(t-i\beta n) -2\cos{\theta})^{\sigma}}+
\sum_{n=1}^{\infty} \frac{e^{-i n\lambda_j}}{(2\cos(t+i\beta n) -2\cos{\theta})^{\sigma}}
\right]
\label{eq:gf_sphere2.2}\,,
\end{multline}
by recognizing the generating functional for the Gegenbauer polynomials $C_l^{(\sigma)}$. The evaluation for $t<0$ is similar. All in all we finally obtain
\begin{multline}
G^{AB}_{T}(t,{\bf y}) = \frac{\Gamma(\sigma)}{4\pi^{\sigma+1}i} 
\sum_{j=1}^N \Psi^{j}_A (\Psi^{j}_B)^\dagger
\\ 
\times\left[
\frac{1}{(4\sin^{2}(\theta/2)-4\sin^{2}(t/2))^{\sigma}}+
\sum_{n\neq 0} \frac{e^{i n \lambda_j}}{(2 \cos(t+i\beta n) - 2 \cos\theta)^\sigma}
\right]
\label{eq:gf_sphere3}\,.
\end{multline}
By subsequently considering the corresponding Wick contractions, we arrive at the following expression for a 
 time-ordered singlet-singlet correlator:
\begin{align}
G(t,{\bf y}) &\equiv \frac{2^{d+2}\pi^{d}}{\Gamma^2(\frac{d-2}{2})} \frac{1}{N}\left\langle \Tr |\Phi^2(t,{\bf y})| \Tr |\Phi^2(0)| \right\rangle \nonumber \\&= 
\sum_{m,n=-\infty}^{\infty} \frac{\sum_{j=1}^N e^{i (m+n) \lambda_j}}{(\cos(t+i\beta m) -  \cos\theta)^\frac{d-2}{2}(\cos(t+i\beta n) -  \cos\theta)^\frac{d-2}{2}}\,.
\label{eq:singsingreallam_app}
\end{align}
The numerical factor has been chosen as to simplify all following expressions, while the factor of $1/N$ guarantees that the two-point function does not scale with $N$.

\section{Fourier space}
\label{app:fourier}

Correlators of glueball operators $\text{Tr}F^2$ in the free limit of ${\cal N} = 4$ SYM were investigated in \cite{HartnollKumar2005} for a flat boundary in mixed $k$-$t$-space as well as in $k$-$\omega$-space. To compare, we can take the planar limit of our expressions in $d=4$ and consider the corresponding Fourier transforms. As we shall see, in the limit of a $\delta$-function eigenvalue distribution, our expressions agree with those of \cite{HartnollKumar2005} up to the different number of derivatives in the definition of the operators. In the second part of this appendix we use the resulting expressions to directly extract the evanescent modes of section \ref{sec:evanescence}.
\subsection{Retarded propagator}
In the planar limit, the retarded propagator Eq.~\eqref{eq:gret} becomes
\begin{equation}
G_R(x,t)  = 4\Theta(t)\left(\text{Im}(- t^2+x^2)^{-2}+2\,\text{Im}(- t^2+x^2)^{-1}\sum_{m=1}^{\infty} \rho_{m}\,\text{Re}\left[((\beta m - i t)^2+x^2)^{-1}\right]\right)\,.
\label{eq:gfret_flat}
\end{equation}
After appropriately introducing $\epsilon$-factors to circumvent the poles, this expression can straightforwardly be transformed to momentum space to read
\begin{equation}
\label{eq:mixed_ret}
G_R(k,t) = \int d^3x\,e^{i {\bf k x}}\,G_R(x,t) = 4\Theta(t)\left(\frac{\cos k t}{2t}+4 \frac{\sin k t}{k} \sum_{m=1}^{\infty}\frac{\rho_m}{4 t^2+m^2\beta^2}\right)\,.
\end{equation}

In the low temperature phase, when $\rho(m) \approx \delta_{m0}$, only the vacuum contributes. Beyond the phase transition, on the other hand, when $\rho(m) \approx 1$, we can carry out the summation to obtain
\begin{equation}
\label{eq:retmixed}
G_R(k,t) =  \frac{4\Theta(t)}{k}\left(\frac{k t \cos k t - \sin k t}{2 t^2}+\frac{\pi \sin k t }{t \beta}\coth\left(\frac{2\pi t}{\beta}\right)\right)\,.
\end{equation}

We can readily transform the above expression to frequency space. We find, up to an overall factor of $4$ due to our normalization in accordance with \cite{HartnollKumar2005},
\begin{equation}
\label{eq:retfourier}
G_R(k,\omega
) = -2+\frac{4\pi i}{k\beta}\log\left(\frac{\Gamma(-\frac{i\beta}{4\pi}(\omega-k))}{\Gamma(-\frac{i\beta}{4\pi}(\omega+k))}\right)-\left(\frac{2 \pi i}{k\beta}-\frac{\omega}{k}\right)\log\frac{\omega+k}{\omega-k}\,.
\end{equation}

\subsection{Evanescence}
In principle, we can extract the evanescent behavior derived in section \ref{sec:evanescence} directly from Eq.~\eqref{eq:retfourier}. However, to allow for a nontrivial eigenvalue distribution, we will instead use Eq.~\eqref{eq:mixed_ret} as our starting point.
Since by definition, $G_R(k,t)$ is real, the imaginary part of $G_R(k,\omega)$ corresponds to the Fourier sine transform of $G_R(k,t)$. At the same time, since the term inside the brackets in Eq.~\eqref{eq:mixed_ret} is odd under reflection of $t$, the domain of integration for the sine transform can be extended to the whole $t$-domain. Consequently, the integral can be directly calculated by contour integration. Explicitly, we have
\begin{equation}
\text{Im }G_R(k,\omega) =2 \int_{-\infty}^\infty \left(\frac{\cos k t}{2t}+4 \frac{\sin k t}{k} \sum_{m=1}^{\infty}\frac{\rho_m}{4 t^2+m^2\beta^2}\right) \sin \omega t\,,
\end{equation}
and thus
\begin{multline}
\text{Im }G_R(k,\omega) =\frac{\pi}{2}\left(\text{sgn}(k+\omega) - \text{sgn}(k-\omega)\right)\\-
\sum_{m=1}^{\infty}\frac{\pi\rho_m}{2km\beta}\left(\Theta(k+\omega)e^{-\frac{m\beta}{2}(k+\omega)}-\Theta(k-\omega)e^{-\frac{m\beta}{2}(k-\omega)}\right.\\\left.+\Theta(-k-\omega)e^{-\frac{m\beta}{2}(-k-\omega)}-\Theta(\omega-k)e^{-\frac{m\beta}{2}(\omega-k)}\right)\,.
\end{multline}
In the evanescent regime, neither the vacuum term nor the two final terms in the above sum contribute. Moreover, the dominant contribution arises for $m = 1$, due to the Boltzmann factors. These considerations imply the final result
\begin{equation}
\text{Im }G_R(k,\omega) \approx \frac{\pi\rho_1}{2k\beta}\left(e^{-\frac{\beta}{2}(k-\omega)}-e^{-\frac{\beta}{2}(k+\omega)}\right) \approx \frac{\pi}{4}\frac{\omega\rho_1}{k}e^{-\frac{\beta}{2}k}\,,
\end{equation}
in agreement with Eq.~\eqref{eq:im_prop_ev_fin} for $d=4$.

\section{Consistency check on bulk and boundary retarded propagator support}\label{sec:supportConsistency}

We can check explicitly how a low temperature bulk propagator which does not vanish inside the light cone is consistent with a boundary correlator without support inside the light cone, as should be the case in even boundary/odd bulk dimension. 
The CFT operator that we are considering has scaling dimension $\Delta = d-2$ for a $d$-dimensional boundary theory. In the dual bulk theory, it is this conformal dimension which determines the analytic properties of the bulk-bulk correlator. In particular, consider its explicit expression at $T = 0$ in global coordinates, reviewed for example in \cite{Nastase2015},
\begin{equation}
G_\Delta(\xi) = C_\Delta\,\xi^\Delta
\, {}_2F_1\left(\frac{\Delta}{2},\frac{\Delta+1}{2},\Delta - \frac{d}{2}+1,\xi^2\right)\,,
\label{eq:bulkbulkprop}
\end{equation}
where $C_\Delta$ is a normalization factor unimportant to our discussion, ${}_2F_1$ is a hypergeometric function and 
\begin{equation}
\xi \equiv \frac{\cos\rho \cos\rho'}{\cos t - \sin \rho \sin \rho' \,\cos\theta}\,,
\end{equation}
with $\rho$ labeling the radial direction
in global coordinates, with the boundary being at $\rho = \pi/2$ and the center of AdS at $\rho = 0$.

For $\Delta = d-2$, the hypergeometric function simplifies and the bulk expression takes on the form
\begin{equation}
G_\Delta(\xi) = C_\Delta\,\xi^{d-2} (1-\xi^2)^\frac{1-d}{2}\,.
\end{equation}
Hence, it picks up an imaginary part for even boundary/odd bulk dimensions\footnote{In particular, this reassuringly implies that for a four-dimensional bulk, where the scalar is conformally coupled, the imaginary part of the propagator is purely due to its singularities.} whenever $\xi > 1$. This imaginary part, however, is confined to the bulk interior, as follows directly from the boundary-boundary correlator, which we obtain as 
\begin{equation}
G(t,\theta) = C_\Delta^{-1}\lim_{\rho \to \frac{\pi}{2}} \lim_{\rho' \to \frac{\pi}{2}} \cos^{-\Delta}\rho\, \cos^{-\Delta}\rho'\, G_\Delta(\xi)=\frac{1}{(\cos t - \cos \theta)^\Delta}\label{eq:bdy-bdy}\,,
\end{equation}
in agreement with Eqs.~\eqref{eq:singsingreallamvec} and \eqref{eq:singsingreallamadj} at $T\to 0$.
As long as the bulk is well described by thermal AdS-space, this behavior persists, as follows from a direct extrapolation of Eq.~\eqref{eq:bulkbulkprop} to finite temperatures.

\bibliographystyle{utphys}
\bibliography{Bibliography}
\end{document}